\documentclass[10pt]{article}
\pdfoutput=1

\usepackage{amssymb}
\usepackage{amsmath}
\usepackage{mathtools}
\usepackage{algpseudocode}
\usepackage{bm}
\usepackage{graphics,epsfig}
\DeclareGraphicsExtensions{.pdf}

\usepackage{url,here}

\usepackage{algorithm}
\allowdisplaybreaks
\usepackage[backref,colorlinks=true]{hyperref}
\usepackage{multirow}

\begin{document}

\newenvironment {proof}{{\noindent\bf Proof.}}{\hfill $\Box$ \medskip}

\newtheorem{theorem}{Theorem}[section]
\newtheorem{lemma}[theorem]{Lemma}
\newtheorem{condition}[theorem]{Condition}
\newtheorem{proposition}[theorem]{Proposition}
\newtheorem{remark}[theorem]{Remark}
\newtheorem{definition}[theorem]{Definition}
\newtheorem{hypothesis}[theorem]{Hypothesis}
\newtheorem{corollary}[theorem]{Corollary}
\newtheorem{example}[theorem]{Example}
\newtheorem{descript}[theorem]{Description}
\newtheorem{assumption}[theorem]{Assumption}
\newcommand{\ag}[1]{{\color{black} #1}}

\def\P{\mathbb{P}}
\def\R{\mathbb{R}}
\def\E{\mathbb{E}}
\def\N{\mathbb{N}}
\def\Z{\mathbb{Z}}

\renewcommand {\theequation}{\arabic{section}.\arabic{equation}}
\def \non{{\nonumber}}
\def \hat{\widehat}
\def \tilde{\widetilde}
\def \bar{\overline}

\def\ind{{\mathchoice {\rm 1\mskip-4mu l} {\rm 1\mskip-4mu l}
{\rm 1\mskip-4.5mu l} {\rm 1\mskip-5mu l}}}
\algnewcommand{\algorithmicgoto}{\textbf{go to step}}%
\algnewcommand{\Goto}[1]{\algorithmicgoto~\ref{#1}}%

\title{\Large\ { \bf A finite state projection algorithm for the stationary solution of the chemical master equation}}

\author{Ankit Gupta, Jan Mikelson and Mustafa Khammash \\
Department of Biosystems Science and Engineering \\ ETH Zurich \\  Mattenstrasse 26 \\ 4058 Basel, Switzerland. 
}
\date{}

\maketitle
\begin{abstract}
The chemical master equation (CME) is frequently used in systems biology to quantify the effects of stochastic fluctuations that arise due to biomolecular species with low copy numbers. The CME is a system of ordinary differential equations that describes the evolution of probability density for each population vector in the state-space of the stochastic reaction dynamics. For many examples of interest, this state-space is infinite, making it difficult to obtain exact solutions of the CME. To deal with this problem, the Finite State Projection (FSP) algorithm was developed by Munsky and Khammash (Jour.\ Chem.\ Phys.\ 2006), to provide approximate solutions to the CME by truncating the state-space. The FSP works well for finite time-periods but it cannot be used for estimating the stationary solutions of CMEs, which are often of interest in systems biology. The aim of this paper is to develop a version of FSP which we refer to as the {\sl stationary} FSP (sFSP) that allows one to obtain accurate approximations of the stationary solutions of a CME by solving a finite linear-algebraic system that yields the stationary distribution of a continuous-time Markov chain over the truncated state-space. We derive bounds for the approximation error incurred by sFSP and we establish that under certain stability conditions, these errors can be made arbitrarily small by appropriately expanding the truncated state-space. We provide several examples to illustrate our sFSP method and demonstrate its efficiency in estimating the stationary distributions. In particular, we show that using a quantized tensor train (QTT) implementation of our sFSP method, problems admitting more than 100 million states can be efficiently solved. 
\end{abstract}


\medskip

\setcounter{equation}{0}

\section{Introduction} \label{sec:intro}
Many intracellular reaction networks consist of biomolecular species that are typically present in low copy-numbers. The reactions involving these species fire \emph{intermittently} at random times, rather than continuously. Hence deterministic descriptions of the reaction dynamics, based on Ordinary Differential Equations (ODEs), become highly inaccurate \cite{Arkin}. It is now well-known that macroscopic properties of the system can be heavily influenced by the \emph{intrinsic noise} or \emph{randomness} that arises due to the random timing of reactions \cite{Elowitz}. Consequently stochastic formulations of the reaction dynamics, based on continuous-time Markov chains (CTMCs), has become a popular approach for studying the effects of intrinsic noise \cite{DASurvey}. In this paper we provide a tool for the analysis of such models.

In the CTMC model of a reaction network, the state at any time is the vector of copy-number counts of all the species. When the number of network species is $d$, the dynamics evolves on a discrete state-space $\mathcal{E}$ which is a subset of the $d$-dimensional nonnegative integer lattice $\N^d_0$ and this subset must be large enough to include all the states that are accessible by the random dynamics. The effects of intrinsic noise on the reaction network are generally studied using the probability distribution $p(t)$ of the random state-vector $X(t)$ at time $t$. It is known that the time-evolution of this probability distribution is given by a system of coupled ODEs, known as the Chemical Master Equation (CME) in the literature (see \eqref{defn:cme}). For each state in $\mathcal{E}$ there is an ODE in the CME that captures the inflow and outflow of probability at that state. If this state-space $\mathcal{E}$ is finite, then the CME is a finite system of linear ODEs which can in principle be solved to yield the probability distribution $p(t)$. However in many examples of biological interest, the state-space $\mathcal{E}$ is infinite, making the CME impossible to solve. A common approach in such cases is to estimate the CME solution by computing the empirical distribution of the samples obtained by simulating the CTMC using Monte Carlo methods such as Gillespie's Stochastic Simulation Algorithm (SSA) \cite{GP}. This simulation-based approach can be very time-consuming and the estimates suffer from \emph{statistical errors} due to finitely many samples being used. In particular the low-probability events are sparsely sampled by Monte Carlo simulations, which can lead to incorrect representations of the CME solution. Such problems can be avoided by using the \emph{Finite State Projection} (FSP) method  developed by Munsky and Khammash \cite{FSP}, that directly solves the CME by truncating the state-space $\mathcal{E}$ to a manageable size (see Section \ref{subsec:fspmethod}). The solution obtained is approximate but FSP provides an iterative way to ensure that the approximation error is within some pre-specified tolerance level.

The truncated state-space needed by FSP to solve the CME accurately is still exorbitantly large for many problems of interest. For example, consider a simple gene-expression network where ten protein species are interacting with each other. Typically each protein in a cell has copy-numbers of the order of several thousands. So even if we have a conservative upper-bound of 1000 on the copy-number of each protein, the size of the state-space required for FSP is of the order of $10^{30}$, which is beyond the computational and storage capacity of modern day computers. This combinatorial explosion in the state-space size is often called the ``curse of dimensionality" and it presents a major challenge in making the CME practically solvable. Several advanced numerical techniques have been developed that address this challenge by adapting the FSP. These techniques include Krylov Subspace approximations \cite{Krylov}, Tensor-Train representations \cite{Kazeev}, and using sparse grids and aggregation methods \cite{SparseGrid1}. Unlike these methods which attempt to solve the exact version of CME, there also exist a body of methods that aim to solve \emph{simplified} versions of CME, which are derived by approximating the CTMC dynamics by a Stochastic Differential Equation (SDE) or a Piecewise-deterministic Markov Process (PDMP) (see \cite{KurtzLLn2,Hellander,Benni}). Such dynamical approximations only hold for finite time-periods, and the assumptions on species copy-numbers and reaction propensities they require, are not always satisfied by networks encountered in systems biology.

For many biological applications, one is interested in the steady-state behavior which is captured by the stationary probability distribution $\pi$ to which the solution $p(t)$ of the CME converges to as $t \to \infty$. For CTMCs whose state-space $\mathcal{E}$ is finite and not \emph{too large}, estimation of the stationary distribution $\pi$ is a simple linear-algebraic problem (see \eqref{defn:stationarydistribution}). However in situations where the state-space is \emph{very large} or infinite, this linear-algebraic problem cannot be practically solved, and we need to estimate $\pi$ by other means. The methods mentioned above for estimating $p(t)$ only work over finite time-intervals and they would generally fail to provide an accurate estimate of the stationary distribution $\pi$. The reason for this failure depends on the method being used. The dynamical approximations based on PDMPs or SDEs introduce an error that can become unbounded in the limit $t \to \infty$, and the Monte Carlo simulation based approach for estimating $\pi$ is highly undesirable due to statistical errors and the computational costs associated with these simulations over large time-intervals. The FSP algorithm also cannot be used for estimating $\pi$ because this method introduces an \emph{absorbing state} to catch all the transitions that leave the truncated state-space (see Figure \ref{figure:lattice}{\bf B}). However in the limit $t \to \infty$, all the probability mass flows into this absorbing state, and so the obtained probability distributions are unable to capture the true stationary distribution $\pi$. We revisit this point later in this section and also explain it in detail in Section \ref{subsec:fspmethod}.

The aim of this paper is to present a FSP-like method for accurately estimating the stationary distribution $\pi$. This method also involves truncating the state-space but rather than solving a linear system of ODEs for probabilities over the truncated state-space (as in FSP), our method estimates the true stationary distribution $\pi$ by computing the stationary distribution of a suitably defined CTMC over the truncated state-space. As the latter step can be accomplished by solving a linear-algebraic system, rather than a system of ODEs, the computational complexity of our method is much lower than that of FSP. Consequently it can be successfully applied on a larger class of networks. We call our method the \emph{stationary Finite State Projection} (or sFSP) algorithm and we provide theoretical arguments to establish its accuracy under certain stability conditions which are usually satisfied by networks in systems and synthetic biology. Even though sFSP can be applied on larger systems than FSP, the combinatorial explosion of state-space sizes still limits the range of applicability of sFSP severely. As was the case with FSP, this issue can be somewhat resolved by adapting sFSP to work with quantized tensor-train (QTT) representations \cite{Kazeev}, sparse grids and aggregation methods \cite{SparseGrid1}. We illustrate this point with a computational example where sFSP is applied to the QTT representation of the CME (see Section \ref{secqttsfsp}). We remark here that QTT representations have already been successfully employed for obtaining approximations of the stationary distribution for reaction networks satisfying certain graph-theoretic criteria \cite{Kazeev2015,ACKProd}. However these criteria are highly-restrictive and it will become evident that the sFSP based approach is more versatile.

We now describe the problem of estimating stationary distributions in more detail. Henceforth let $|A|$ denote the size of any set $A$, and let ${\bf 0}$ and ${\bf 1}$ denote the vector of all zeros and all ones respectively\footnote{The dimension of these vectors will be clear from the context.}. The stochastic model of a reaction network (see Section \ref{sec:stochasticmodel}) represents the dynamics as a CTMC over a discrete state-space $\mathcal{E} \subset \N^d_0$. Such a CTMC can be described by its $|\mathcal{E}| \times |\mathcal{E}|$ transition rate matrix $Q$ (see \cite{Norris}), whose diagonal entries are non-positive, off-diagonal entries are non-negative and all the rows sum to zero. The stationary distribution for this CTMC can be described by a non-negative vector\footnote{Throughout the paper we assume that vector and matrix indices start from $0$ rather than $1$.} $\pi = (\pi_0,\pi_1,\dots)$, which is in the left null-space of transition rate matrix $Q$, i.e. 
\begin{align}
\label{defn:stationarydistribution}
Q^T \pi = {\bf 0},
\end{align}
and its components sum to $1$ (i.e. ${\bf 1}^T \pi = \sum_{i} \pi_i = 1$). Such a stationary distribution may not be unique and if $|\mathcal{E}| = \infty$ then it may not even exist (see \cite{Norris}). In our recent work, we have dealt extensively with the issue of computationally verifying the existence and uniqueness of the stationary distribution corresponding to the CTMC models for a large class of biomolecular reaction networks (see \cite{GuptaIrred} and \cite{GuptaPLOS}). Assuming that the existence and uniqueness of the stationary distribution $\pi$ has been ascertained for the network, our aim here is to estimate $\pi$ numerically. We are primarily interested in situations where $\mathcal{E}$ is infinite, and so the direct computation of $\pi$ using \eqref{defn:stationarydistribution} is computationally impossible.

It is natural to try to estimate $\pi$ by solving a finite, truncated version of the linear-algebraic system \eqref{defn:stationarydistribution}. This truncated version can be obtained by first identifying a truncated state-space and then \emph{projecting} the CTMC dynamics on this truncated state-space. Thereafter the stationary distribution for the projected CTMC, found by solving the corresponding linear-algebraic system of the form \eqref{defn:stationarydistribution}, serves as an estimate of the true stationary distribution $\pi$. An important issue that arises here is how to handle the outgoing transitions from the truncated state-space, so that the obtained estimate of $\pi$ is accurate. In the FSP approach \cite{FSP}, these outgoing transitions are preserved but their target states are collapsed into a single absorbing state (see Figure \ref{figure:lattice}{\bf B}). This leads to the ``probability leakage" problem which can be managed over finite time-intervals but not in the asymptotic $t \to \infty$ regime. This problem manifests itself in the fact that the only stationary distribution for the projected CTMC would be the one that puts all the probability-mass at the absorbing state. Obviously this does not capture the true stationary distribution and hence modifications to the FSP approach are necessary to circumvent the probability-leakage problem. One such modification that has been tried is motivated by the use of ``reflected" boundary conditions in the study of Fokker-Planck equations \cite{Kazeev2015}. In this approach all the outgoing transitions from the truncated state-space are simply eliminated by setting their propensities to zero. It has been shown that this reflected version of FSP yields accurate estimates of the stationary distribution for some reaction network examples \cite{Gauckler,Kazeev2015}. However there is no theoretical guarantee that this approach will work well in general.

The method sFSP that we present in this paper modifies the FSP in another way. It preserves the outgoing transitions from the truncated state-space, but rather than channeling them to an absorbing state (as in FSP), it redirects them to a designated state \emph{within the truncated state-space} (see Figure \ref{figure:lattice}{\bf C}). This modification is simple to implement and its appealing feature is that for a wide range of biomolecular reaction networks, we can theoretically guarantee that the stationary distribution of the projected CTMC converges to the actual stationary distribution $\pi$ as the truncated state-space expands to the full state-space $\mathcal{E}$. Moreover we derive bounds for the approximation error incurred by this approach, in terms of the \emph{outflow rate} of all the outgoing transitions evaluated at the estimated stationary distribution (see Theorem \ref{thm:mainresult}). These results provide the theoretical basis for our method which expands the truncated state-space iteratively to recover a ``good" approximation of $\pi$. Note that our approach for estimating the stationary distribution is very different from the \emph{stochastic complementation} approach proposed in \cite{Verena}. This complementation approach is generally difficult to implement for infinite state-space CTMCs and it only yields the conditional stationary distribution which can then be used to derive upper and lower bounds for the true stationary probabilities. However such bounds are not guaranteed to be close to each other. In contrast, our method allows one to estimate the true stationary probabilities directly.

For our method sFSP to work we require that the original CTMC representing the reaction network satisfies a couple of stability conditions. The first condition is that the state-space $\mathcal{E}$ needs to be \emph{irreducible} i.e. all the states in $\mathcal{E}$ must be accessible from each other via a sequence of positive-propensity reactions. The second condition is a Foster-Lyapunov criterion (see \cite{Meyn}) which ensures that the original CTMC is \emph{exponentially ergodic} i.e.\ the solution $p(t)$ of the CME converges to the stationary distribution $\pi$ exponentially fast. We elaborate these stability conditions in Section \ref{section:mainconvergenceresult} and there we also explain how these conditions can be easily checked for a wide range of networks arising in systems and synthetic biology, using the computational procedures developed in our recent papers \cite{GuptaIrred} and \cite{GuptaPLOS}. This makes the proposed sFSP method broadly applicable and of interest to the growing community of researchers working with stochastic models of biomolecular reaction networks.

This paper is organized as follows. In Section \ref{sec:prelim} we describe the stochastic model and the original FSP method \cite{FSP}. In Section \ref{sec:sFSP} we present and mathematically analyze our \emph{stationary Finite State Projection} (or sFSP) algorithm. A simple implementation of sFSP is presented in Section \ref{secsimplesfsp} while its QTT implementation is presented in Section \ref{secqttsfsp}. These sections also include the computational examples that illustrate the respective implementations. Finally in Section \ref{sec:conclusion} we conclude and discuss directions for future research.

\section{Preliminaries} \label{sec:prelim}

\subsection{The stochastic model} \label{sec:stochasticmodel}

We now formally describe the CTMC model of a reaction network. Suppose this network has $d$ species, called $\mathbf{X}_1,\dots,\mathbf{X}_d$, and $K$ reactions of the form
\begin{align}
\label{formofthereaction}
\sum_{i=1}^d \nu_{ik} \mathbf{X}_i \longrightarrow \sum_{i=1}^d \nu'_{ik} \mathbf{X}_i.
\end{align} 
Here $\nu_{ik}$ and $\nu'_{ik}$ are nonnegative integers denoting the number of molecules of species $\mathbf{X}_i$ that are consumed and produced by the $k$-th reaction. The state of the system at any time is the vector $x = (x_1,\dots,x_d) \in \N^d_0$ of molecular counts of all the $d$ species. When the $k$-th reaction fires, the state is displaced by the \emph{stoichiometric} vector $\zeta_k \in \Z^d$ whose $i$-th component is $\zeta_{ik} = (\nu'_{ik}  - \nu_{ik} )$. At any state $x$, the rate of the $k$-th reaction is $\lambda_k(x)$, where $\lambda_k : \N^d_0 \to [0,\infty)$ is the propensity function for this reaction. Commonly \emph{mass action kinetics} (see \cite{DASurvey}) is assumed, where each $\lambda_k$ is given by
\begin{align}
\label{massactionkinetics}
\lambda_k(x_1,\dots,x_d) = \theta_k \prod_{i =1}^d \frac{ x_i(x_i-1)\dots (x_i - \nu_{ik} +1 ) }{  \nu_{ik} ! },
\end{align}
with the positive parameter $\theta_k$ being the associated rate constant. We model the reaction dynamics as a CTMC which jumps from state $x$ after a random waiting time which is exponentially distributed with rate $\lambda_0(x) := \sum_{k=1}^K \lambda_k(x)$, and this jump is in direction $\zeta_k$ with probability $\lambda_k(x) / \lambda_0(x) $. Formally this CTMC can be specified by its generator\footnote{The generator of a Markov process is an operator which specifies the rate of change of the probability distribution of the process (see Chapter 4 in \cite{EK} for details).} $\mathbb{Q}$ defined as
\begin{align}
\label{defn_gen}
\mathbb{Q} f (x)  = \sum_{k=1}^K \lambda_k(x ) \left( f(x+\zeta_k) - f(x) \right),
\end{align}
where $f$ is any bounded real-valued function on $\N^d_0$.

From now on we suppose that there is a nonempty state-space $\mathcal{E} \subset \N^d_0$ on which the CTMC evolves i.e. 
\begin{align}
\label{defn_state_space}
\textnormal{ for each } x \in \mathcal{E} \textnormal{ and } k=1,\dots,K, \quad \textnormal{if} \ \lambda_k(x) > 0 \textnormal{ then } \ (x+\zeta_k) \in \mathcal{E}.
\end{align}
In other words, if at state $x \in \mathcal{E}$, reaction $k$ has a positive probability of firing then the resulting state $(x+\zeta_k)$ must also be in $\mathcal{E}$. As $\mathcal{E}$ is at most countable, it can be \emph{enumerated}. This means that we can find a one-to-one and onto map $\phi$ from $\mathcal{E}$ to the set $\{0,1,\dots, |\mathcal{E}| -1 \}$. Once such an enumeration is fixed, the set $\mathcal{E}$ can be expressed as $\mathcal{E} = \{x_0,x_1,\dots\}$, where $x_i = \phi^{-1}(i)$. Then the CTMC generator $\mathbb{Q}$ can be expressed as the transition rate matrix $Q = [Q_{ij}]$ given by\footnote{Here we assume for convenience that all stoichiometry vectors ($ \zeta_k$-s) are distinct.  } 
\begin{align*}
Q_{ij} = \left\{  
\begin{array}{cc}
- \sum_{k=1}^K \lambda_k(x_i)  & \textnormal{ if } i = j  \\
 \lambda_k(x_i) &  \textnormal{ if }  x_j = x_i + \zeta_k \textnormal{ for some } k \\
 0 & \textnormal{ otherwise}.  
\end{array} \right.
\end{align*}
Let $( X(t) )_{t  \geq 0}$ be the CTMC with this transition rate matrix and some initial state $X(0)  \in \mathcal{E}$. For any state $x \in \mathcal{E}$, let
\begin{align}
\label{transition_probabilities}
p(t,x) = \P\left( X(t) = x \right)
\end{align}
be the probability that the CTMC is in state $x$ at time $t$. These probabilities evolve in time according to the Chemical Master Equation (CME) given by
\begin{align}
\label{defn:cme}
\frac{ d p(t,x) }{dt} = & \sum_{k=1}^K  p(t, x -\zeta_k) \lambda_k(x - \zeta_k) -p(t, x ) \sum_{k=1}^K \lambda_k(x),
\end{align}
for each $x \in \mathcal{E}$. Note that this system has as many ODEs as the number of elements in the state-space $\mathcal{E}$, which is generally infinite or very large.

Let $p(t)$ be the probability distribution defined by
\begin{align}
\label{defn_prob_measure}
p(t,A) = \sum_{x \in A } p(t,x)
\end{align}
for any $A \subset \mathcal{E}$. The vectorized form of $p(t)$ w.r.t. enumeration $\phi$ is simply given by $$p(t) = ( p(t,x_0),   p(t,x_1),  p(t,x_2),\dots  )$$ and using this form we can express the CME as
\begin{align}
\label{cme_equiv}
\frac{d p}{d t} = Q^T p(t).
\end{align}
If the number of states in $\mathcal{E}$ is finite, then this first-order system can in principle be solved by exponentiating the matrix $Q^T$, i.e.\ the solution is given by
\begin{align*}
p(t) = \exp( Q^T t )p(0)  \quad \textnormal{for any} \quad t  \geq 0,
\end{align*}
where $p(0)$ is the vectorized form of the probability distribution of the initial state $X(0)$. However this approach is infeasible for large state-spaces and in such cases, the Finite State Projection (FSP) method \cite{FSP} can be used to approximately solve the CME (see Section \ref{subsec:fspmethod}).

In many biological applications, rather than the finite-time behavior, one is interested in the properties of the system after it has settled down, or in other words, the CTMC $(X(t))_{t  \geq 0}$ has reached a \emph{steady-state} which is characterized by a stationary distribution $\pi$ satisfying \eqref{defn:stationarydistribution}, that is essentially a fixed-point for the CME \eqref{cme_equiv}. We say that the CTMC $(X(t))_{t  \geq 0}$ is \emph{ergodic} if this fixed-point is unique and \emph{globally attracting} in the sense that for any initial probability distribution $p(0)$, the solution $p(t)$ of \eqref{cme_equiv} satisfies
\begin{align}
\label{defn:ergodicity}
\lim_{ t \to \infty } \| p(t)  - \pi \|_{  \ell_1}  = 0,
\end{align}
where $ \| p(t)  - \pi \|_{  \ell_1}=\sum_{x \in \mathcal{E} } | p(t,x)  - \pi(x) | $ denotes the $\ell_1$-distance\footnote{Generally ergodicity is defined using the \emph{total variation} distance between probability distributions. However for a discrete state-space $\mathcal{E}$ the total variation distance among probability distributions is exactly half of the distance computed using the $\ell_1$ norm. So we work with the $\ell_1$ norm in this paper.} between probability measures $p(t)$ and $\pi$. Furthermore the CTMC is called \emph{exponentially ergodic} if the convergence in \eqref{defn:ergodicity} is exponentially fast, i.e. there exist positive constants $C$ and $\rho$ such that for any $t > 0$   
\begin{align*}
 \| p(t)  - \pi \|_{  \ell_1}   \leq C e^{ - \rho t }.
\end{align*}
Here the constant $C$ may depend on the initial distribution $p(0)$ but the constant $\rho$ does not (see \cite{Meyn} for example).

\subsection{The Finite State Projection Algorithm} \label{subsec:fspmethod}

In the FSP method, approximate solutions of the CME \eqref{cme_equiv} are obtained by restricting it to a truncated state-space. Suppose this truncated subset is given by a finite set $\mathcal{E}_n \subset \mathcal{E} $ of size $n = | \mathcal{E}_n |$. Using the same enumeration $\phi$ as in Section \ref{sec:stochasticmodel}, we can express the set $\mathcal{E}_n$ as $\mathcal{E}_n = \{x_{j_1},x_{j_2},\dots,x_{j_n}  \}$. Letting $Q_n$ to be the matrix formed by the rows and columns of matrix $Q$ in the set $J_n:=\{j_1,\dots,j_n\}$, we approximate \eqref{cme_equiv} by the $n$-dimensional linear system   
\begin{align}
\label{cme_fsp}
\frac{d p_n}{d t} = Q_{ n }^T p_n(t).
\end{align}
The solution of this system is simply $p_n(t) = \exp( Q_n^T t )p_n(0)$, where $p_n(0)$ is the $n \times 1$ containing the components of vector $p(0)$ in the set $J_n$.

Let ${\bf 1}$ be the $n$-dimensional vector of all ones. We assume that the initial state $X(0)$ can only take values in $\mathcal{E}_n$ and so ${\bf 1}^T  p_n(0) = 1$. It is easy to check that all the rows of matrix $Q_n$ have a non-positive sum, which implies that for any $t \geq 0$
\begin{align*}
\epsilon_n(t):= 1 -  {\bf 1}^T  p_n(t) =1 -  {\bf 1}^T  \exp( Q_n^T t )p_n(0)  \geq 0.
\end{align*}
Results in \cite{FSP} show that $\epsilon_n(t)$ quantifies the ``error" between the actual solution of CME $p(t)$ and its approximation $p_n(t)$. For any fixed $t$, this error $\epsilon_n(t)$ decreases monotonically with increasing values of $n$. Moreover as $n \to \infty$ and the truncated state-space $\mathcal{E}_n$ approaches the full state-space $\mathcal{E}$, we have $\epsilon_n(t) \to 0$. In the FSP algorithm of \cite{FSP}, the final time $t_f$ is fixed and the system \eqref{cme_fsp} is solved in the time-interval $[0,t_f]$ with some truncated state-space $\mathcal{E}_n$. Thereafter the error $\epsilon_n(t_f)$ is evaluated and if this value is below some pre-specified tolerance level $\epsilon$, then the algorithm is terminated. Otherwise the truncated state-space is expanded to include more states and the same is process is repeated. After finitely many such iterations, the truncated state-space becomes large enough to ensure that the tolerance criterion is met.

Another way to formulate the FSP method is to consider the \emph{projected} CTMC over the state-space $\tilde{\mathcal{E}}_n = \mathcal{E}_n \cup \{x_A\}$, whose transitions among the states in $ \mathcal{E}_n$ are same as the original CTMC, but any outgoing transitions from the set $ \mathcal{E}_n$ are \emph{absorbed} in the state $x_A$ (see Figure \ref{figure:lattice}{\bf B}), which serves as a proxy for all states in the set $\mathcal{E}^c_n = \{ x \in \mathcal{E} : x \notin \mathcal{E}_n  \}$. Enumerating the elements of $\tilde{\mathcal{E}}_n$ as $\tilde{\mathcal{E}}_n = \{ x_{j_1},\dots,x_{j_n},x_A \}$, the $(n+1) \times (n+1)$ transition rate matrix $\tilde{Q}_n$ for this projected CTMC is given by
\begin{align}
\label{defn:qntilde}
\tilde{Q}_n = \left[  
\begin{array}{cc}
Q_n & c_n \\
{\bf 0} & 0  
\end{array}
\right],
\end{align}
where $c_n$ is the $n$-dimensional column vector whose $i$-th component is
\begin{align}
\label{defn_outflowrates}
c_{n,i} =  \sum_{k =1, (x_{j_i} +\zeta_k) \notin \mathcal{E}_n }^K \lambda_k(x_{j_i}).
\end{align}
This choice of $c_n$ ensures that all the rows of matrix $\tilde{Q}_n $ sum to $0$ and hence $\tilde{Q}_n$ is a valid transition rate matrix. Let $\tilde{p}_n(t)$ be the solution of the CME \eqref{cme_equiv} corresponding to rate matrix $\tilde{Q}_n $ and with initial value $\tilde{p}_n(0) =( p_n(0) , 0 ) $. Then it can be shown that for any $t \geq 0$ we can express $\tilde{p}_n(t)$ as
\begin{align*}
\tilde{p}_n(t) = (p_n(t), \epsilon_n(t)),
\end{align*}
which proves that the FSP approximation error $ \epsilon_n(t)$ at time $t$ is exactly the amount of probability-mass that has been absorbed by the state $x_A$ in the time-interval $[0,t]$.

One can show that typically for any fixed truncated state-space $\mathcal{E}_n$, we would have $\epsilon_n(t) \to 1$ as $t \to \infty$, which says that eventually all the probability mass gets absorbed by the state $x_A$. Therefore even if the original CTMC is ergodic and the solution $p(t)$ of the CME \eqref{defn:cme} converges to $\pi$ as $t \to \infty$, the approximate solution $p_n(t)$ obtained by solving the FSP system \eqref{cme_fsp}, will {\bf not} be close to $\pi$ for large times and in fact $p_n(t)$ converges to a vector of all zeros at $t \to \infty$. This is also evident from the stationary distribution $\tilde{\pi}_n$ that can be computed by finding a non-zero solution to the linear-algebraic system \eqref{defn:stationarydistribution} corresponding to the matrix $\tilde{Q}_n$ (see \eqref{defn:qntilde}). This would yield a stationary distribution of the form $\tilde{\pi}_n = ( {\bf 0} , 1)$, which assigns all the mass to the absorbing state $x_A$ and hence $\tilde{\pi}_n $ cannot be close to $\pi$. This shows that the FSP approach is not conducive for the estimation of stationary distribution $\pi$.

\section{The stationary FSP method} \label{sec:sFSP}

\begin{figure*}[t]
\centering
\includegraphics[width=\textwidth]{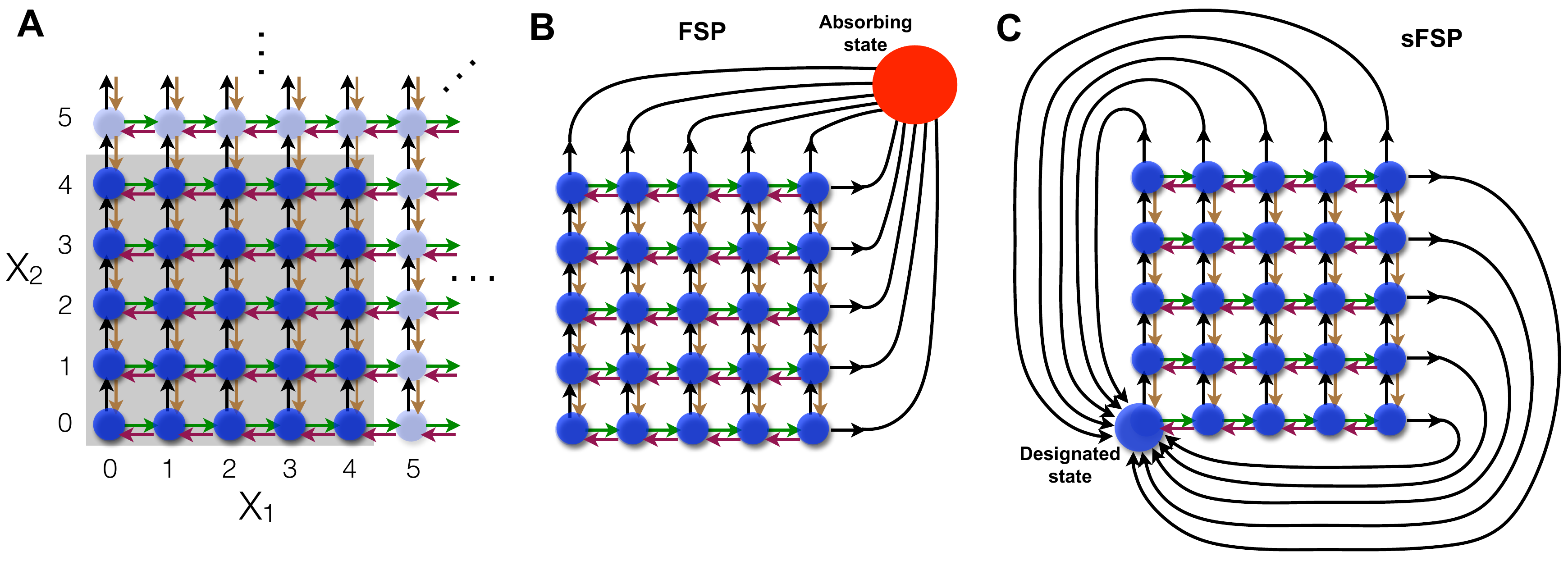}
\caption{Consider the state-space $\mathcal{E} = \N^2_0$ of a two-species network shown in panel {\bf A} along with a finite truncation of this set (see the \emph{Grey box}). Panels {\bf B} and {\bf C} depict how the reaction dynamics is projected onto this truncated state-space in the FSP method \cite{FSP} and in the sFSP method developed in this paper. While in FSP the outgoing transitions are directed to an \emph{absorbing} state (see panel {\bf B}), in sFSP these outgoing transitions are redirected to some \emph{designated state} within the truncated state-space (see panel {\bf C}). }
\label{figure:lattice}
\end{figure*}

In this section we present our method sFSP for estimating the stationary distribution $\pi$ for the CTMC model of a reaction network. This is accomplished by constructing a \emph{projected} CTMC over the truncated state-space and computing the stationary distribution of this new CTMC. Keeping the same notation as in Section \ref{subsec:fspmethod}, this projected CTMC over the truncated state-space $\mathcal{E}_n = \{x_{j_1},\dots,x_{j_n}\}  \subset \mathcal{E}$ is constructed by redirecting the transitions that leave $\mathcal{E}_n$ to some designated state $x \in \mathcal{E}_n $ (see Figure \ref{figure:lattice}{\bf C}). Let the $n \times n$ matrix $Q_n$ and the $n \times 1$ vector $c_n$ be as in Section \ref{subsec:fspmethod}. Then the $n \times n$ transition rate matrix $\bar{Q}_n$ for this CTMC is simply given by
\begin{align}
\label{defn:qnbar}
\bar{Q}_n = Q_n + c_n b_l,
\end{align}
where $l$ corresponds to the \emph{address} of the designated state (i.e. $x_{j_l} = x$) and $b_l$ is the $1 \times n$ vector whose $l$-th component is $1$ and the rest are all zeros. Essentially, $\bar{Q}_n$ is formed by adding the non-negative vector $c_n$ to the $l$-th column of matrix $Q_n$. All the rows of matrix $\bar{Q}_n $ sum to $0$ and hence $\bar{Q}_n$ is a valid transition rate matrix and so our projected CTMC is well-defined. Our method sFSP estimates the stationary distribution $\pi$ by computing the finite-dimensional stationary distribution $\bar{\pi}_n$ for the projected CTMC with transition rate matrix $\bar{Q}_n$. Using $\bar{\pi}_n$, we can also compute the overall \emph{outflow rate} at the estimated stationary distribution by
\begin{align}
\label{defn:outflowrate}
r^{(n)}_{ \textnormal{out} } = c^T_n \bar{\pi}_n.
\end{align}
This quantity will play a key role in bounding the sFSP approximation error whose direct computation is impossible.

\subsection{Analysis of sFSP} \label{section:mainconvergenceresult}

 The aim of this section is to demonstrate that under certain conditions, that are commonly satisfied by biological reaction networks, the sFSP approximation error can be made arbitrarily small by picking a truncated state-space $\mathcal{E}_n$, that is \emph{large enough}. Moreover it is possible to check if $\mathcal{E}_n$ is large enough by computing a \emph{convergence factor} which is defined by suitably scaling the outflow rate $r^{(n)}_{ \textnormal{out} }$. The main results of this section are collected in Theorem \ref{thm:mainresult} and they provide the theoretical basis for our sFSP method.

Before we present our result we need to discuss some preliminary concepts. The state-space $\mathcal{E}$ of the original CTMC $(X(t))_{t \geq 0}$ is called \emph{irreducible} if this CTMC has a positive probability of reaching any state in $\mathcal{E}$ from any other state in $\mathcal{E}$, in a finite time. More formally, the state-space $\mathcal{E}$ is irreducible, if for any $x,y \in \mathcal{E}$ we have $ \P( X(t) = y \vert X(0) = x  ) > 0$ for some $t > 0$. In our setting of reaction networks, this is equivalent to saying that between any two states $x,y \in  \mathcal{E}$ there exists a sequence of positive-propensity reactions $k_1,\dots,k_n$ that takes the dynamics from $x$ to $y$. For this to hold we must have $y = x+ \sum_{i=1}^n \zeta_{k_i}$ and at each intermediate state $z_j = (x + \sum_{i=1}^{j-1} \zeta_{k_i})$ the next reaction in the sequence ($k_{j}$) has a positive propensity of firing ($\lambda_{k_j} (z_j) >0 $). When only finitely many states are accessible by the reaction dynamics, irreducible state-spaces can be easily found by manipulating the transition rate matrix $Q$ (see \cite{Kemeny}). However when infinitely many states are accessible, finding irreducible state-spaces within the infinite lattice becomes a complicated task. In a recent work \cite{GuptaIrred} we address this challenge and develop a computational procedure that can find all the irreducible state-spaces for a large class of biological reaction networks. In particular, for most networks of interest each irreducible state-space has the form\footnote{To obtain this form relabeling of species may be required.}
\begin{align}
\label{irredstatespaceform}
\mathcal{E} = \mathcal{E}_b \times \N^{d_f}_0,
\end{align}
where $\mathcal{E}_b$ is a finite set in $\N^{d_b}$, and $d_b, d_f$ are non-negative integers summing up to the total number of species $d$. Here $\mathcal{E}_b$ contains the dynamics of $d_b$ \emph{bounded} species whose copy-numbers are required to satisfy a positive mass-conversation relation. A typical example is a gene-expression network where the gene of interest has many (say $d_b$) activity modes. To represent the dynamics we need to represent each such mode by a different network species, but all these species will be \emph{bounded} and their copy-numbers will evolve in a finite set $\mathcal{E}_b$, because the gene of interest has a fixed copy-number (see the Pap-Switch example in Section \ref{sec:examples} for instance). The species that are not \emph{bounded} are \emph{free}\footnote{Apart from \emph{free} and \emph{bounded} species, there may also exist another type of species, called \emph{restricted} species, whose dynamics essentially \emph{mimics} the dynamics of \emph{free} species according to some affine map. However these \emph{restricted} species can be easily eliminated to obtain a dynamically equivalent network and hence we ignore such species here (see \cite{GuptaIrred} for more details).} to have any copy-number and hence the state-space for their dynamics is taken to be the full non-negative integer orthant $\N^{d_f}_0$.

Note that the property of ergodicity (see Section \ref{sec:stochasticmodel}) will obviously fail if there do not exist any stationary distributions or there exist more than one stationary distributions for the CTMC $(X(t))_{t \geq 0}$. If the state-space $\mathcal{E}$ is finite, then its irreducibility is sufficient to guarantee that the stationary distribution exists uniquely and the CTMC is exponentially ergodic (see \cite{Kemeny}). However when $\mathcal{E}$ is infinite, its irreducibility can only guarantee the uniqueness of a stationary distribution but the existence of this distribution must be checked by other means, for example, using the results in \cite{MeynandTweedieBook} and \cite{Meyn}. In particular Theorem 7.1 in \cite{Meyn} guarantees the existence of a stationary distribution along with exponential ergodicity, if we can construct a function $V : \mathcal{E} \to [1,\infty)$ which is \emph{norm-like} (i.e. $V(x) \to \infty$ as $\|x\| \to \infty$) and for some $C_1,C_2 >0$, the following holds for all $x \in \mathcal{E}$:
\begin{align}
\label{negativedriftcondition}
\mathbb{Q} V(x)  \leq C_1 - C_2 V(x),
\end{align}
where $\mathbb{Q}$ is the CTMC generator given by \eqref{defn_gen}. This condition is called the Foster-Lyapunov criterion in the literature and it describes the tendency of the CTMC to experience a \emph{drift} towards some finite set in the state-space with a force that is proportional to the distance from this finite set, measured according to $V$. In \cite{GuptaPLOS} it is shown that for many biomolecular reaction networks, a linear Foster-Lyapunov function
\begin{align}
\label{linearlyapunov}
V(x) =  1 + \langle v, x \rangle,
\end{align}
satisfying \eqref{negativedriftcondition} can be constructed. Here $v \in \R^d$ is a positive vector which is chosen using simple Linear Programming and $\langle \cdot,\cdot \rangle$ denotes the standard inner product in $\R^d$. Observe that for the linear function $V(x)$ \eqref{linearlyapunov}, the drift condition \eqref{negativedriftcondition} is simply
\begin{align}
\label{negativedriftcondition2}
\sum_{k=1}^K \lambda_k(x) \langle v, \zeta_k \rangle   \leq C_1 -C_2 (1+ \langle v, x \rangle)   \quad  \textnormal{ for all } \quad x \in \mathcal{E}.
\end{align}
As demonstrated in \cite{GuptaPLOS}, often for biological reaction networks the vector $v$ can be chosen in such a way that along with this drift condition, the following \emph{diffusivity} condition is also satisfied - for some $C_3,C_4 > 0$
\begin{align}
\label{diffusivitycondition}
\sum_{k=1}^K \lambda_k(x) \langle v, \zeta_k \rangle^2   \leq C_3  +  C_4 (1 + \langle v, x \rangle)   \quad  \textnormal{ for all } \quad x \in \mathcal{E}.
\end{align}  
When \eqref{negativedriftcondition2} and \eqref{diffusivitycondition} hold simultaneously, then in addition to exponential ergodicity, one can also guarantee other desirable properties like finiteness of all statistical moments of the stationary distribution $\pi$ and convergence of all the moments of the CTMC to their steady-state values as time approaches infinity (see Theorem 5 in \cite{GuptaPLOS}).

To study the sFSP approximation error we need to work with the norm prescribed by the Foster-Lyapunov function $V$. For any signed measure $\mu$ on $\mathcal{E}$, this norm is given by
\begin{align*}
\|\mu\|_V = \sum_{x \in \mathcal{E}} | \mu(x) | V(x).
\end{align*}
Note that this norm is \emph{tighter} than the $\ell_1$ norm because $\| \mu \|_V \geq \|\mu\|_{\ell_1}$ as $V\geq 1$. Let $\mathcal{B} ( \mathcal{E}_n )$ denote the \emph{boundary} of the truncated state-space $\mathcal{E}_n$, which includes all those states in $\mathcal{E}_n$ for which there exists a positive-propensity reaction that takes the dynamics outside $\mathcal{E}_n$, i.e.
\begin{align}
\label{defn_bdry}
\mathcal{B} ( \mathcal{E}_n ) = \{ x \in \mathcal{E}_n :  \lambda_k(x) > 0  \textnormal{ and } (x + \zeta_k) \notin  \mathcal{E}_n \textnormal{ for some } k=1,\dots,K \}.
\end{align}
Based on the outflow rate $r^{(n)}_{ \textnormal{out} }$ given by \eqref{defn:outflowrate}, we define the \emph{convergence factor} as
\begin{align}
\label{defn_conv_factor}
\gamma^{(n)}_V =  r^{(n)}_{ \textnormal{out} }  \| \mathcal{E}_n \|_{V},
\end{align}
where \begin{align}
\label{defn_emaxv}
 \| \mathcal{E}_n \|_{V} = V(x_{\ell}) +  \max_{ x \in \mathcal{B}( \mathcal{E}_n ) } V(x)   
\end{align} 
and $x_\ell$ is the designated state. Our next result will show that the convergence factor $\gamma^{(n)}_V$ is a useful diagnostic tool to assess the approximation error $\| \pi - \bar{\pi}_n \|_{V}$ of sFSP. Note that unlike the approximation error, $\gamma^{(n)}_V$ can be explicitly computed from the sFSP output $\bar{\pi}_n$ if the Foster-Lyapunov function $V$ is known. In situations where $V$ is unknown, the definition of $\gamma^{(n)}_V$ can often be suitably modified to preserve its diagnostic purpose (see Remark \ref{thm:remark2}).

We now come to the main result of our paper.

\begin{theorem}
\label{thm:mainresult}
Suppose that state-space $\mathcal{E}$ is irreducible for the original CTMC with transition rate matrix $Q$, and there exists a Foster-Lyapunov function $V: \mathcal{E} \to [1,\infty)$ satisfying \eqref{negativedriftcondition}. Also assume that $\{ \mathcal{E}_n : n =1,2,\dots \}$ is a sequence of finite sets that is increasing (i.e. $ \mathcal{E}_{n_1} \subset  \mathcal{E}_{n_2}$ if $n_1 < n_2$) and that covers the full state-space $\mathcal{E}$ in the limit $n \to \infty$. Fix a designated state $x_\ell \in \mathcal{E}_1$ and let $\bar{Q}_n$ be the transition rate matrix of our projected CTMC with state-space $\mathcal{E}_n$, defined according to \eqref{defn:qnbar}. Then we have the following:
\begin{itemize}
\item[(A)] The stationary distribution $ \bar{ \pi }_n$ for the projected CTMC exists uniquely.
\item[(B)] As $n \to \infty$, $ \bar{ \pi }_n$ converges to the stationary distribution $\pi$ for the original CTMC, in the $\ell_1$ metric, i.e.
\begin{align}
\label{main_prop_limit}
\lim_{n \to \infty} \|  \pi -  \bar{\pi}_n \|_{ \ell_1 } = 0.
\end{align}
\item[(C)] There exists a positive constant $M$ such that for any $n$
\begin{align}
\label{defn_error_measure}
 \|  \pi -  \bar{\pi}_n \|_{ V} \leq M   \gamma^{(n)}_V,
\end{align}
where $ \gamma^{(n)}_V$ is the convergence factor defined by \eqref{defn_conv_factor}.
\item[(D)] Suppose that the Foster-Lyapunov function $V$ has the linear form \eqref{linearlyapunov} and the positive vector $v$ is such that both  \eqref{negativedriftcondition2} and \eqref{diffusivitycondition} are satisfied. Furthermore assume that the sequence of sets $\{ \mathcal{E}_n \}$ grows uniformly w.r.t. function $V$ which means that for some constant $\theta  \in (0,1)$ we have
\begin{align}
\label{unif_growth_cond}
 \min_{ x \in \mathcal{B}( \mathcal{E}_n ) }  V(x)  \geq \theta   \max_{ x \in  \mathcal{B}( \mathcal{E}_n )  }  V(x) \quad \textnormal{for all} \quad n =1,2,\dots,
\end{align}
where $\mathcal{B}( \mathcal{E}_n )$ is the \emph{boundary} of $ \mathcal{E}_n $ defined by \eqref{defn_bdry}. Then there exists a constant $M' > 0$ for which the converse of \eqref{defn_error_measure} also holds, i.e. for each $n$
\begin{align}
\label{defn_error_measure_conv}
 \|  \pi -  \bar{\pi}_n \|_{ V } \geq M'   \gamma^{(n)}_V.
\end{align}
Furthermore, the convergence factor $ \gamma^{(n)}_V$ converges to $0$ as $n  \to \infty$.
\end{itemize}
\end{theorem}

\begin{remark}
\label{thm:remark1}
It will become evident from the proof that if the Foster-Lyapunov function $V$ and constants $C_1,C_2$ in \eqref{negativedriftcondition} are known, then a constant $M$ satisfying part (C) can be explicitly computed using the results in Meyn and Tweedie \cite{meyn1994computable}. Hence part (C) provides a computable upper-bound for the approximation error $ \|  \pi -  \bar{\pi}_n \|_{ V }$. Similarly the constant $M'$ satisfying part (D) may be explicitly computed from constants $C_1,\dots,C_4$ in \eqref{negativedriftcondition2} and \eqref{diffusivitycondition}, and the constant $\theta$ that appears in \eqref{unif_growth_cond}. The tightness of the error bounds obtained from these explicitly computable constants remains to be investigated. Nevertheless parts (C) and (D) are useful in demonstrating that up to a constant, the magnitude of the uncomputable approximation error $ \|  \pi -  \bar{\pi}_n \|_{ V}$ can be assessed by computing the convergence factor $\gamma^{(n)}_V$. In other words, if $\gamma^{(n)}_V \leq \epsilon$ then $ \|  \pi -  \bar{\pi}_n \|_{ V} \leq M \epsilon$, and similarly if $\gamma^{(n)}_V \geq \epsilon$ then $ \|  \pi -  \bar{\pi}_n \|_{ V } \geq M' \epsilon$, where $M$ and $M'$ are the optimal constants for which parts (C) and (D) hold.
\end{remark}
\begin{remark}
\label{thm:remark2}
Note that computation of the convergence factor $\gamma^{(n)}_V$ \eqref{defn_conv_factor} requires knowledge of the Foster-Lyapunov function $V$ which is undesirable from the point of view of applications. However it is possible to circumvent this problem, if one has information about the form of $V$ and the shape of finite sets $\{ \mathcal{E}_n \}$. For this one needs to pick a sequence $\{ \beta_n\}$ such that for some constants $ \alpha, \alpha' > 0$ 
\begin{align*}
\frac{1}{\alpha} \| \mathcal{E}_n \|_V \leq \beta_n \leq \frac{1}{ \alpha'}  \| \mathcal{E}_n \|_V, 
\end{align*}
holds for each $n$, with $\| \mathcal{E}_n \|_V$ defined by \eqref{defn_emaxv}. Then one can define the convergence factor as
\begin{align}
\label{new_conv_factor}
\gamma_n =  r^{(n)}_{ \textnormal{out} } \beta_n,
\end{align}
with the outflow rate $ r^{(n)}_{ \textnormal{out} } $ given by \eqref{defn:outflowrate}, and parts (C) and (D) will hold with the substitutions, $\gamma^{(n)}_V \to \gamma_n$, $M \to M \alpha$ and $M' \to M' \alpha'$. For example, if $V$ has the linear form \eqref{linearlyapunov}, then one can define $\beta_n$ in the same way as $\| \mathcal{E}_n \|_V$ but with $V(x)$ replaced by any norm $\|x\|$ on $\R^d$.
\end{remark}

\begin{proof}
We start by proving part (A). The stationary distribution $\bar{\pi}_n$ for the projected CTMC certainly exists because the transition rate matrix $\bar{Q}_n$ is finite (see \cite{Kemeny}). This stationary distribution can be found by solving the linear-algebraic system \eqref{defn:stationarydistribution} with transition-rate matrix $\bar{Q}_n$. We now prove by contradiction the uniqueness of this stationary distribution. Suppose that this uniqueness does not hold. Then there would exist at least two disjoint non-empty irreducible state-spaces (say $A$ and $B$) for the projected CTMC within the state-space $\mathcal{E}_n$. This implies that if the projected CTMC starts in set $A$ then it remains in this set for all times, and there is a positive probability for this CTMC to reach any state in $A$ from any other state in $A$ in a finite time. The same holds true for set $B$. Certainly one of these sets, say $A$, will not contain the designated state $x_{\ell}$ but this leads to a contradiction due to the following reasons. Since the state-space $\mathcal{E}$ is irreducible for the original CTMC, there exists a sequence of reactions $k_1,\dots,k_m$ that takes the original CTMC from any state $x \in A$ to the designated state $x_{\ell}$ with a positive probability. If all the intermediate states that arise in this reaction path (recall $z_j$-s from above) lie within the set $\mathcal{E}_n$, then the same sequence of reactions will also take the projected CTMC from state $x \in A$ to state $x_{j_l}$, which is a contradiction because $A$ is an irreducible state-space not containing $x_{j_l}$. On the other hand if one of the intermediate states lies outside $\mathcal{E}_n$, then the last reaction, say $k_q$, in the sequence that takes the dynamics outside $\mathcal{E}_n$ will be redirected to the designated state $x_{\ell}$ in the projected CTMC and hence again we have a contradiction because $k_1,\dots,k_q$ is a positive-probability sequence of reactions that takes the projected CTMC from state $x \in A$ to state $x_{j_l}\notin A$. Therefore the stationary distribution $\bar{\pi}_n$ for the projected CTMC is unique, and this completes the proof of part (A).  

We now prove part (B). Clearly the assertion of part (B) is trivial when the full state-space $\mathcal{E}$ is finite and so we assume that $\mathcal{E}$ is infinite from now on. Let $\{ \mathcal{E}_n \}$ be a sequence of sets as stated in the proposition and let $\phi : \mathcal{E} \to \N_0$ be an enumeration of $\mathcal{E}$ satisfying 
\begin{align}
\label{enum_cond}
\phi(x) \in \{0,1,\dots, | \mathcal{E}_n| -1\} \quad \textnormal{for each} \quad x \in \mathcal{E}_n \quad \textnormal{and} \quad n =1,2,\dots.
\end{align}
Such an enumeration exists because $\{ \mathcal{E}_n \}$ is an increasing sequence of sets that cover the set $\mathcal{E}$ in the limit $n \to \infty$. Note that as each $\mathcal{E}_n$ is a finite set, condition \eqref{enum_cond} ensures that
\begin{align}
\label{phi_limits}
\lim_{ \| x \| \to \infty } \phi(x) = \infty \qquad \textnormal{and} \qquad \lim_{i \to \infty} \|  \phi^{-1}(i) \| = \infty.
\end{align}
Now consider the $\N_0$-valued, one-dimensional process $( Y(t) )_{t  \geq 0} $ given by $Y(t) = \phi( X(t) )$ for each $t \geq 0$, where $( X(t) )_{t  \geq 0} $ is the original CTMC with transition rate matrix $Q$ and generator $\mathbb{Q}$ (see \eqref{defn_gen}). As $\phi$ is a one-to-one and onto map, the process $( Y(t) )_{t  \geq 0} $ is also a CTMC and its generator is given by
\begin{align*}
\hat{ \mathbb{Q} } g( i) = \mathbb{Q} f( \phi^{-1} (i) ),
\end{align*}
where $g$ is a bounded real-valued function on $\N_0$ and $f$ is the bounded real-valued function on $\N^d_0$ defined by $f(x) = g( \phi(x) )$. 

Irreducibility of state-space $\mathcal{E}$ for $( X(t) )_{t  \geq 0} $ implies the irreducibility of state-space $\N_0$ for $( Y(t) )_{t  \geq 0} $. Let $V: \N^d_0 \to [0,\infty)$ be the norm-like Foster-Lyapunov function satisfying \eqref{negativedriftcondition} and define the function $\hat{V} : \N_0 \to [0,\infty)$ by $\hat{V}(i) = V( \phi^{-1}(i) )$. Then using \eqref{phi_limits} and \eqref{negativedriftcondition} we can deduce that $\hat{V}$ is a norm-like function satisfying
\begin{align*}
\hat{ \mathbb{Q} } \hat{V}( i) = \mathbb{Q} V( \phi^{-1}(i)  ) \leq C_1 -C_2 V( \phi^{-1}(i)  )  = C_1 - C_2  \hat{V}( i). 
\end{align*}
Therefore $ \hat{V}$ is a Foster-Lyapunov function for CTMC $( Y(t) )_{t  \geq 0} $ with generator $\hat{ \mathbb{Q} }$ and hence this CTMC is exponentially ergodic due to Theorem 7.1 in \cite{Meyn}. Let $\hat{\pi}$ and $\hat{\pi}_n$ be the probability distributions on $\N_0$ and $\{0,1,\dots, | \mathcal{E}_n |-1\}$ defined by
\begin{align*}
\hat{\pi}(i) = \pi( \phi^{-1}(i) ) \quad \textnormal{and} \quad  \hat{\pi}_n(i) = \bar{\pi}_n( \phi^{-1}(i) ). 
\end{align*}
Then $\hat{\pi}$ is the stationary distribution for the CTMC $( Y(t) )_{t  \geq 0} $ and $ \hat{\pi}_n$ is the stationary distribution of this CTMC projected onto the finite state-space $\{0,1,\dots, | \mathcal{E}_n |-1\}$ by redirecting all the outgoing transitions to the designated state $\phi(x_\ell)$. Theorem 3.3 in \cite{Hart} proves
\begin{align*}
\lim_{n \to \infty} \| \hat{\pi}  -  \hat{\pi}_n \|_{ \ell_1 } = 0,
\end{align*}
using resolvent forms (see \eqref{resol_defn}). This limit is equivalent to \eqref{main_prop_limit} and this proves part (B).

We will now prove part (C). Without loss of generality we can assume that $\mathcal{E}_n = \{0,1,\dots,n-1\}$. Define an infinite vector
\begin{align}
\label{defn_vartheta}
\vartheta_n = Q^T \left[ 
\begin{array}{c}
\bar{\pi}_n \\
{\bf 0} 
\end{array} \right] = \left[ 
\begin{array}{c}
\vartheta_1 \\
\vartheta_2 
\end{array} \right],
\end{align}
whose first $n$ elements are $\vartheta_1 = Q_n^T \bar{\pi}_n$, where $Q_n$ denotes the $n \times n$ northwest sub-matrix of $Q$. Recall that matrix $\bar{Q}_n$ is given by \eqref{defn:qnbar} and the outflow rate $r^{ (n) }_{\textnormal{out}}$ is defined by \eqref{defn:outflowrate}. As $\bar{Q}^T_n \bar{\pi}_n = {\bf 0}$ we can write $\vartheta_1 $ as
\begin{align*}
\vartheta_1 = Q^T_n \bar{\pi}_n - \bar{Q}^T_n \bar{\pi}_n = (Q_n - \bar{Q}_n)^T  \bar{\pi}_n = - b_l^T c^T_n  \bar{\pi}_n = - b_l^T r^{ (n) }_{\textnormal{out}},
\end{align*}
which shows that the $n \times 1$ vector $\vartheta_1$ has only one non-zero entry which is equal to $-r^{ (n) }_{\textnormal{out}}$ and it is at the position corresponding to the designated state $x_\ell$. Since $Q {\bf 1} = {\bf 0}$ we have ${\bf 1}^T \vartheta_n = {\bf 0}$ which implies that
\begin{align*}
{\bf 1}^T \vartheta_2 = - {\bf 1}^T \vartheta_1  = c^T_n  \bar{\pi}_n = r^{ (n) }_{\textnormal{out}}.
\end{align*}
One can check that all entries of the infinite vector $\vartheta_2$ are non-negative and only those entries are non-zero that correspond to the states in the boundary set $\mathcal{B}( \mathcal{E}_n )$ (see \eqref{defn_bdry}) of $\mathcal{E}_n$. Therefore, viewing $\vartheta_n$ as a signed measure over $\mathcal{E}$, we can express it as
\begin{align}
\label{vartheta_exp1}
\vartheta_n = r^{(n)}_{ \textnormal{out} } ( \mu_2  - \mu_1), 
\end{align}
where $\mu_1$ and $\mu_2$ are probability measures on $\mathcal{E}$, supported on $\{x_\ell \}$ and $\mathcal{B}( \mathcal{E}_n )$ respectively. With a slight abuse of notation, we will denote the vector-version of $\mu_i$ also as $\mu_i$.

Define the sFSP approximation error in vector form as
\begin{align*}
\epsilon_n = \left(   \pi   - \left[ 
\begin{array}{c}
\bar{\pi}_n \\
{\bf 0} 
\end{array} \right] \right),
\end{align*}
and since $Q^T \pi = {\bf 0}$ we get the following equation from \eqref{defn_vartheta}
\begin{align}
\label{linearsystems_error_basis}
Q^T \epsilon_n =  - \vartheta_n.
\end{align}
One can verify that $\epsilon_n $ is the unique solution of this linear system with the constraint $\langle {\bf 1}, \epsilon_n \rangle = 0$. For any $\beta > 0$, let $R_\beta$ denote the $\beta$-resolvent matrix corresponding to the transition rate matrix $Q$. It is defined by
\begin{align}
\label{resol_defn}
R_\beta = \beta ( \beta {\bf I} - Q )^{-1},
\end{align}
where ${\bf I}$ denotes the identity matrix. It is known (see \cite{Hart}) that $R_\beta$ is a positive matrix satisfying $R_{\beta} {\bf 1} = {\bf 1}$,  $\pi^T R_{\beta} = \pi^T$ and
\begin{align}
\label{resol:forward_backward}
R_\beta = {\bf I} + \beta^{-1} Q R_\beta = {\bf I} + \beta^{-1}R_\beta Q.
\end{align}
One can regard $R_\beta$ as the transition matrix of a discrete-time Markov chain over $\mathcal{E} = \{x_0,x_1,\dots\}$ whose unique stationary distribution is $\pi$.

Expressing the Foster-Lyapunov function $V$ as the vector $V = ( V(x_0), V(x_1),\dots )$ we can write the drift condition \eqref{negativedriftcondition} as
\begin{align*}
Q V \leq C_1 {\bf 1} - C_2 V.
\end{align*}
This relation along with \eqref{resol:forward_backward} and the positivity of $R_\beta$ implies 
\begin{align*}
R_\beta V &= ({\bf I} + \beta^{-1}R_\beta Q) V = V +  \beta^{-1}R_\beta Q V  \leq  V +  \beta^{-1}R_\beta (C_1 {\bf 1} - C_2 V )  = V + \frac{C_1}{\beta} {\bf 1}  - \frac{C_2}{\beta}R_\beta V.
\end{align*}
Letting  $\lambda  = ( 1 + C_2/\beta )^{-1}$ and $C = C_1/( \lambda \beta)$ we obtain
\begin{align*}
R_\beta V \leq \lambda V + C {\bf 1}.
\end{align*}
Note that $\lambda \in  (0,1)$. Theorem 6.1 in \cite{meyn1994computable} shows that we can explicitly compute constants $C' > 0$ and $\rho \in (0,1)$, such that for any probability distribution $\mu$ over $\mathcal{E}$ we have
\begin{align}
\label{resolventbound1}
\| \mu^T R^m_\beta  - \pi^T \|_V \leq C' \|  \mu \|_V \rho^m,
\end{align}
where $R^m_\beta$ denotes the $m$-th power of the matrix $R_\beta$. Transposing \eqref{linearsystems_error_basis}, multiplying both sides by $R_\beta$ and using \eqref{resol:forward_backward} and \eqref{vartheta_exp1} we get
\begin{align*}
\beta \epsilon^T_n ( R_\beta - {\bf I} ) =  \epsilon_n^T Q R_\beta = - \vartheta_n^T R_\beta = r^{ (n) }_{ \textnormal{out} } (\mu^T_1 - \mu^T_2) R_\beta.
\end{align*}
One can write $ \epsilon_n$ as
\begin{align}
\label{error_resol_formula}
\epsilon_n = \frac{r^{ (n) }_{ \textnormal{out} } }{ \beta }( \epsilon_1  - \epsilon_2 ),
\end{align}
where $\epsilon_j$ is the solution to
\begin{align*}
\epsilon_j^T (R_\beta -  {\bf I}) = ( \mu_j  - \pi )^T R_\beta = \mu^T_j R_{\beta} - \pi^T,
\end{align*}
for $j=1,2$. This solution can be expressed as
\begin{align*}
\epsilon_j^T = - ( \mu_j  - \pi )^T \sum_{m=1}^\infty  R^m_\beta =   - \sum_{m=1}^\infty ( \mu^T_j  R^m_\beta - \pi^T ),
\end{align*}
and using \eqref{resolventbound1} we get
\begin{align*}
\| \epsilon_j \|_V \leq  \sum_{m=1}^\infty \| \mu^T_j  R^m_\beta - \pi^T \|_V \leq C'  \| \mu_j \|_V  \sum_{m=1}^\infty \rho^m = \frac{ C' \|\mu_j  \|_V \rho  }{1 - \rho}.
\end{align*}
Therefore
\begin{align*}
\|  \epsilon_n   \|_V \leq \frac{r^{ (n) }_{ \textnormal{out} } }{ \beta }( \| \epsilon_1\|_V  + \|  \epsilon_2\|_V ) \leq M  r^{ (n) }_{ \textnormal{out} } ( \| \mu_1\|_V  + \|  \mu_2\|_V ) 
\end{align*}
where $M  =C' \beta^{-1} \rho ( 1 - \rho )^{-1}$. As $\mu_1$ and $\mu_2$ are probability distributions supported on $\{x_\ell \}$ and $\mathcal{B}( \mathcal{E}_n )$, we have $ ( \| \mu_1\|_V  + \|  \mu_2\|_V )  \leq  \| \mathcal{E}_n \|_{V} $ (see \eqref{defn_emaxv}). This proves part (C) of the theorem.

We now prove part (D). Here we assume that the Foster-Lyapunov function $V$ has the linear form \eqref{linearlyapunov} and both \eqref{negativedriftcondition2} and \eqref{diffusivitycondition} are satisfied. Note that by rescaling the positive vector $v$ in \eqref{linearlyapunov} if necessary, we can assume that
\begin{align*}
| \langle v,  \zeta_k  \rangle | \leq  \langle v,  \zeta_k  \rangle^2 \quad \textnormal{for each} \quad k=1,\dots,K.
\end{align*}
As $\mathbb{Q} V(x) =\sum_{k=1}^K \lambda_k(x) \langle v,\zeta_k \rangle $ from condition \eqref{negativedriftcondition2} we obtain
\begin{align}
\label{partd_reln2}
| \mathbb{Q} V(x) |  \leq \sum_{k=1}^K \lambda_k(x) | \langle v,\zeta_k \rangle | \leq  \sum_{k=1}^K \lambda_k(x)\langle v,  \zeta_k  \rangle^2 \leq C_3 + C_4 V(x),
\end{align}
for each $x \in \mathcal{E}$. Transposing \eqref{linearsystems_error_basis}, multiplying both sides by vector $V$ and taking absolute values we get
\begin{align}
\label{partd_rel0}
| \epsilon^T_n  Q V |  = | \vartheta^T_n  V |.
\end{align}
Using \eqref{partd_reln2} we can upper-bound the l.h.s. as
\begin{align}
\label{partd_reln1}
| \epsilon^T_n  Q V | = | \langle \epsilon_n, Q V \rangle  | \leq  \langle |\epsilon_n| , |Q V| \rangle \leq C_3 \|\epsilon_n\|_{\ell_1} + C_4 \| \epsilon_n \|_V.
\end{align}
Since $\vartheta_n$ is given by \eqref{vartheta_exp1}, with $\mu_1$ and $\mu_2$ being probability distributions supported on $\{x_\ell \}$ and $\mathcal{B}( \mathcal{E}_n )$ respectively, we can lower-bound the r.h.s. of \eqref{partd_rel0} as
\begin{align*}
 | \vartheta^T_n  V | = r^{(n)}_{ \textnormal{out} }\left(  \mu^T_2 V - \mu^T_1 V \right) \geq    r^{(n)}_{ \textnormal{out} } \left( \min_{x \in  \mathcal{B}( \mathcal{E}_n )} V(x) - V(x_\ell) \right).
\end{align*}
The uniform growth condition \eqref{unif_growth_cond}, along with the fact that $V(x_\ell)$ does not depend on $n$, ensures that there exists a positive constant $\theta'$ such that 
\begin{align*}
 \min_{x \in  \mathcal{B}( \mathcal{E}_n )} V(x)  \geq \theta' \| \mathcal{E}_n \|_{V} + V(x_\ell), 
\end{align*}
for each $n$, and hence obtain the lower-bound
\begin{align*}
 | \vartheta^T_n  V |  \geq  \theta'  r^{(n)}_{ \textnormal{out} }\| \mathcal{E}_n \|_{V}.
\end{align*}
This relation along with \eqref{partd_rel0} and \eqref{partd_reln1} yield
\begin{align*}
r^{(n)}_{ \textnormal{out} }\| \mathcal{E}_n \|_{V} \leq \frac{C_3}{ \theta' } \|\epsilon_n\|_{\ell_1} + \frac{C_4}{ \theta' } \| \epsilon_n \|_V,
\end{align*}
which is sufficient to prove \eqref{defn_error_measure_conv} as $\|\epsilon_n\|_{\ell_1} \leq \|\epsilon_n\|_{V}$.

We now prove the second assertion of part (D), i.e. $ \gamma^{(n)}_V \to 0$ as $n  \to \infty$. For this we first demonstrate that the square of the linear Foster-Lyapunov $V$ will also satisfy the drift condition \eqref{negativedriftcondition}. To see this note that for any $x \in \mathcal{E}$
\begin{align*}
\mathbb{Q} V^2(x) &= \sum_{k=1}^K \lambda_k(x) \left( V^2(x +\zeta_k) - V^2(x)  \right) \\
&=  \sum_{k=1}^K \lambda_k(x) \left( V(x +\zeta_k) - V(x)  \right)^2 + 2 V(x) \mathbb{Q} V(x) \\
& =  \sum_{k=1}^K \lambda_k(x) \langle v,  \zeta_k  \rangle^2 + 2 V(x) \mathbb{Q} V(x).
\end{align*}
Using \eqref{negativedriftcondition2} and \eqref{diffusivitycondition} we obtain
\begin{align*}
\mathbb{Q} V^2(x) \leq C_3 + (C_4 + 2 C_1) V(x) - 2 C_2 V^2(x).
\end{align*}
As $V(x)$ is a semi-norm, the quadratic term will dominate the linear term for all $x$ outside some compact set and hence the drift condition \eqref{negativedriftcondition} will be satisfied by function $V^2(x)$ for some constants $\hat{C}_1,\hat{C}_2>0$. This drift condition also ensures that (see \cite{Meyn}) there exists a constant $L$ such that $$\sum_{x \in \mathcal{E} } | \pi(x) - \bar{\pi}_n(x) | V^2(x)  \leq L$$ for each $n$. Now using Cauchy-Schwarz inequality we get
\begin{align*}
 \| \epsilon_n \|^2_V = \left(  \sum_{x \in \mathcal{E} } | \pi(x) - \bar{\pi}_n(x) | V(x) \right)^2 &\leq  \left(   \sum_{x \in \mathcal{E} }  | \pi(x) - \bar{\pi}_n(x) |  \right)   \left(   \sum_{x \in \mathcal{E} } | \pi(x) - \bar{\pi}_n(x) |V^2(x) \right) \\
 & \leq  \| \epsilon_n \|_{ \ell_1 }  L.
\end{align*}
As $n \to \infty$, part (B) shows that $ \| \epsilon_n \|_{ \ell_1 } \to 0$ and hence $ \| \epsilon_n \|_V  \to 0$ as well. Now \eqref{defn_error_measure_conv}  proves that $ \gamma^{(n)}_V \to 0$ and this concludes the proof of the theorem. 

 \end{proof}

\subsection{The sFSP Algorithm}

Theorem \ref{thm:mainresult} proves that under certain conditions, the sFSP approximation error, measured in a certain norm, converges to $0$ as $n \to \infty$ and the truncated state-space $\mathcal{E}_n$ expands to the fully state-space $\mathcal{E}$. Moreover for any $\mathcal{E}_n$ the magnitude of the approximation error can be judged by computing the convergence factor $\gamma_n$ defined according to \eqref{new_conv_factor} with the sequence $\{\beta_n\}$ chosen as in Remark \ref{thm:remark2}. These results form the basis of our \emph{stationary Finite State Projection} (sFSP) algorithm, that is presented as Algorithm \ref{algo_sfsp}. This algorithm takes as input a $d$-species reaction network $\mathcal{R}$, specified as a set of $K$ reactions with propensity functions $\lambda_1,\dots,\lambda_K$ and stoichiometric vectors $\zeta_1,\dots,\zeta_K$. It is required that the CTMC describing the reaction kinetics admits a Foster-Lyapunov function satisfying \eqref{negativedriftcondition} and its state-space $\mathcal{E}$ is irreducible. These conditions can be checked using the results in \cite{GuptaIrred} and \cite{GuptaPLOS} as discussed before.

Algorithm \ref{algo_sfsp} starts by picking an increasing sequence of finite state-space truncations $\{ \mathcal{E}_i : i =1,2,\dots \}$ as in Theorem \ref{thm:mainresult}, a sequence $\{ \beta_i : i =1,2,\dots \}$ as in Remark \ref{thm:remark2}, and a designated state $x_\ell \in \mathcal{E}_1$. Thereafter for each iteration cycle $i$, the transition rate matrix $\bar{Q}_i$ for the projected CTMC over the truncated state-space $\mathcal{E}_i$ is constructed and its stationary distribution $\bar{\pi}_i$ is found by solving the linear-algebraic system \eqref{defn:stationarydistribution} for matrix $\bar{Q}_i$. Next the outflow rate $ r^{(i)}_{ \textnormal{out} }$ and the convergence factor $\gamma_i =  r^{(i)}_{ \textnormal{out} } \beta_i$ are computed. If this convergence factor is below an acceptable threshold level $\epsilon$ (chosen in step 4 of Algorithm \ref{algo_sfsp}), then sFSP terminates after returning $\bar{\pi}_i$ as the estimate of the true stationary distribution $\pi$. Otherwise if $\gamma_i  \geq \epsilon$, then the algorithm goes into the new iteration cycle with the expanded truncated state-space $\mathcal{E}_{i+1}$.

\begin{algorithm}[H]  
\caption{ Provides an estimate of the stationary distribution $\pi$ for the reaction network $\mathcal{R}$ involving $d$ species.}      
 \label{algo_sfsp}
 \begin{algorithmic}[1]
\Require The CTMC for network $\mathcal{R}$ admits a Foster-Lyapunov function $V$ satisfying \eqref{negativedriftcondition} and its state-space $\mathcal{E}$ is irreducible.
\Function{sFSP}{$ \mathcal{R}, \mathcal{E}$}  
\State  Pick an increasing sequence of finite sets $\{ \mathcal{E}_i : i =1,2,\dots \}$ that covers the full state-space $\mathcal{E}$ in the limit $i \to \infty$. Also pick a sequence $\{ \beta_i : i =1,2,\dots \}$ as in Remark \ref{thm:remark2}.
\State Select a designated state $x_\ell \in  \mathcal{E}_1 $.
\State Initialize the counter $i = 1$ and specify the termination condition through a small positive parameter $\epsilon$.
\State Construct the transition rate matrix $\bar{Q}_i $ according to \eqref{defn:qnbar} for the projected CTMC over the truncated state-space $\mathcal{E}_i$ with designated state $x_\ell$. \label{marker1}
\State Solve the linear-algebraic system \eqref{defn:stationarydistribution} for matrix $\bar{Q}_i$ to obtain $\bar{\pi}_i$.
\State Normalize $\bar{\pi}_i$ so that its component-sum is $1$ and hence it is the stationary distribution for the projected CTMC with transition rate matrix $\bar{Q}_i$.
\State Compute the outflow rate $ r^{(i)}_{ \textnormal{out} }$ according to \eqref{defn:outflowrate} and evaluate the convergence factor $\gamma_i =  r^{(i)}_{ \textnormal{out} } \beta_i$.
\If{ $\gamma_i < \epsilon$ }
\State \Return $ \bar{\pi}_i $ as the estimate of the stationary distribution and terminate.
\EndIf
\State Set $i = i+1$ and \Goto{marker1} 
\EndFunction
\end{algorithmic}
\end{algorithm}

\section{sFSP Algorithm: Simple Implementation} \label{secsimplesfsp}

In this section we present the simple implementation of sFSP akin to to the classical FSP \cite{FSP}, where the multi-dimensional state-space is explicitly enumerated, and accordingly the transition rate matrix for the projected CTMC is constructed and its stationary distribution vector is computed. The performance of sFSP depends crucially on the choice of finite state-space truncations $\mathcal{E}_i$-s and their enumerating functions $\phi_i$-s. We now discuss these choices for our implementation of sFSP.

\subsection{State-space enumeration and truncation}\label{sec:enum}

\begin{figure*}[t]
\centering
\includegraphics[width=\textwidth]{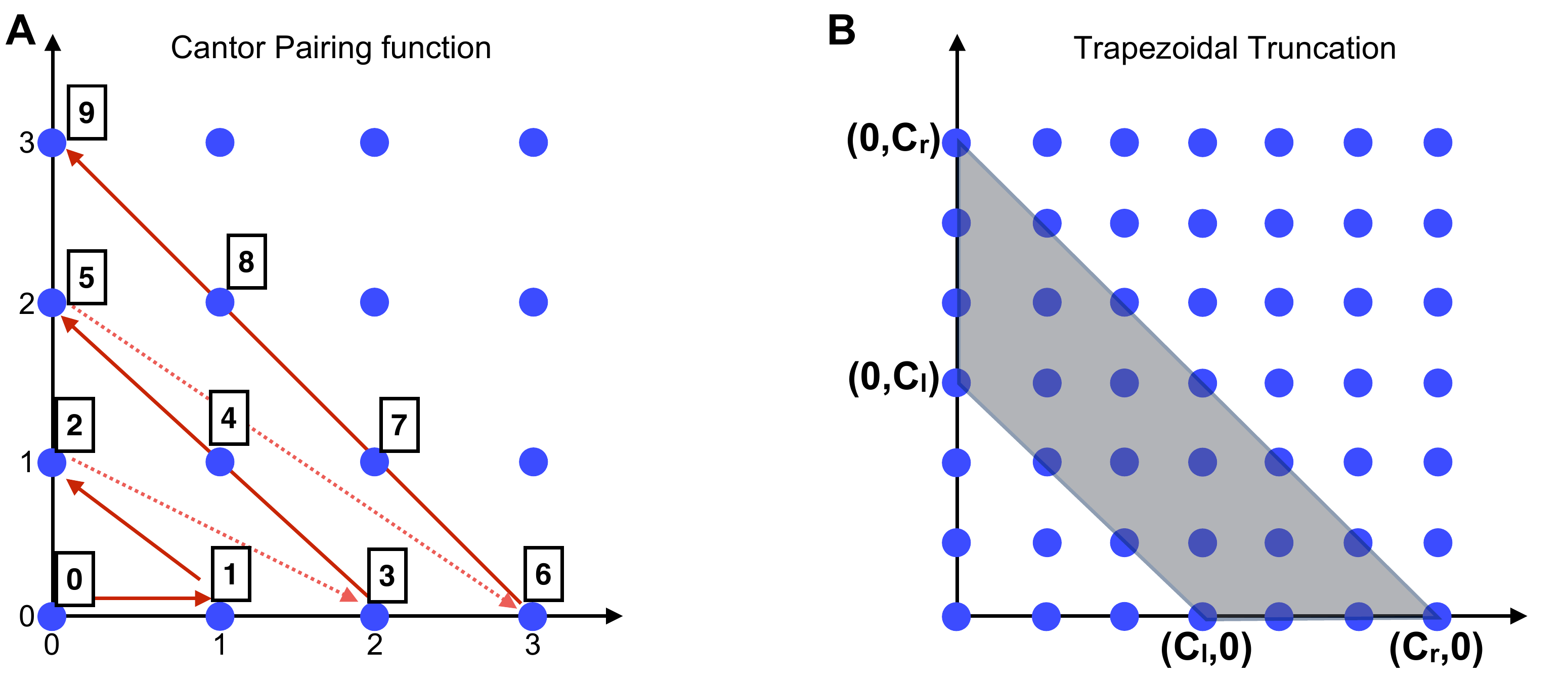}
\caption{Panel {\bf A} depicts the Cantor pairing function as a bijection from $\N^2_0$ to $\N_0$. This map is constructed by moving along the anti-diagonal $x_1+x_2 = k$ sequentially until the state $(0,k)$ is reached (see the \emph{solid red lines}). As this point the map jumps to the next anti-diagonal $x_1+x_2 = (k+1)$ at state $(k+1,0)$ (see the \emph{dotted red lines}) and enumeration process continues. Panel {\bf B} shows a trapezoidal truncation \eqref{defn_section} in two-dimensions (see the \emph{Grey shaded area}). Note that only those states are included in this set whose component-sum $x_1+x_2$ is between the bounds $C_l$ and $C_r$.}
\label{figure:enumeration}
\end{figure*}

The basic ingredient of our state-space enumeration strategy is the \emph{Cantor Pairing} function (see \cite{CantorPairing}) which is the bijective map between $\N^2_0$ to $\N_0$ defined by
\begin{align}
\label{defn_cantor_map}
\Phi_2(x_1,x_2) =  \frac{1}{2}(x_1+ x_2)(x_1 + x_2 + 1) + x_2.
\end{align}
Under this bijection, the elements in $\N^2_0$ are mapped to $\N_0$ by moving along the \emph{anti-diagonals}, which are the straight lines given by $x_1 +x_2 = k$ (see Figure \ref{figure:enumeration}{\bf A}). This map is easy to invert and for any $z \in \N_0$, $(x_1,x_2) = \Phi^{-1}_2(z)$ can be computed as $x_1 = v - x_2$ and $x_2 = z - v(v+1)/2$, where 
 \begin{align*}
v = \left\lfloor \frac{\sqrt{8z+1} -1}{2} \right\rfloor.
\end{align*} 
Henceforth we define $\Phi_1$ as the identity map on $\N_0$. By composition, one can extend the Cantor function to obtain a bijection from $\N^n_0$ to $\N_0$ for any positive integer $n$. Such a bijective map $\Phi_n$ can be defined recursively as
\begin{align*}
\Phi_n( x_1,\dots,x_n ) = \Phi_2( \Phi_{n-1}( x_1,\dots,x_{n-1} ) ,x_n ).
\end{align*}
Similarly the inverse $\Phi^{-1}_n: \N_0 \to \N^n_0$ of this map can also be defined recursively as
 \begin{align*}
\Phi^{-1}_n( z) = ( \Phi^{-1}_{n-1}( z_1 ) ,  z_2 )
\end{align*}
where $(z_1,z_2) = \Phi^{-1}_2(z)$.

Consider the situation where the irreducible state-space $\mathcal{E}$ has the form \eqref{irredstatespaceform} with $d_b = 0$ and $d =d_f$. In this case, $\mathcal{E}$ is just the $d$-dimensional non-negative integer orthant $\N^d_0$ and we enumerate it using the Cantor function $\Phi_d$. An explicit formula for $\Phi_d$ can be obtained (see \cite{CantorPairing}) as
\begin{align*}
\Phi_d(x_1,\dots,x_d) = {x_1+\dots + x_d + d - 1  \choose d} +  {x_2+\dots + x_d + d - 2  \choose d - 1} +\dots + {x_d  \choose 1},
\end{align*}
where ${n \choose k} = \frac{n!}{k! (n-k)!}$ denotes the binomial coefficient. This formula shows that for any $C_l,C_r \in \N_0$ with $C_l \leq C_r$, the following set  
\begin{align}
\label{defn_section}
\mathcal{T}(C_l,C_r) = \{ x  \in \N^{d}_0 :   \Phi_d(C_l, {\bf 0} )  \leq \Phi_d(x) \leq  \Phi_d( {\bf 0},C_r) \},
\end{align}
is non-empty, and we call this set a \emph{trapezoidal truncation} of $\N^d_0$ with left cut-off point $C_l$ and right cut-off point $C_r$. For $d=2$, we plot such a set in Figure \ref{figure:enumeration}{\bf B} and it simply consists of all the states $(x_1,x_2)$ whose component-sum $x_1+x_2$ is between $C_l$ and $C_r$. This may not be exactly true is higher-dimensions ($d >2$) but still one can think of a trapezoidal truncation as the set of states whose component-sum is within certain bounds. Note that in our setting of reaction networks, the component-sum of a state represents the total molecular count of all the species. In many biomolecular reaction networks this total molecular count is within certain tight bounds even though each species can individually have high copy-number variation. This is mainly because the species are often in competition with each other, through mechanisms such as mutual repression or interconversion, which ensures that the total molecular count is tightly regulated. This property makes trapezoidal truncations very appealing for our purpose of estimating stationary distributions. This point is nicely illustrated by the Toggle-Switch example considered in Section \ref{sec:ts}.

We now consider the situation where the irreducible state-space $\mathcal{E}$ has the form \eqref{irredstatespaceform} for some finite non-empty set $\mathcal{E}_b \subset \N^{d_b}_0$. Let $N_b = | \mathcal{E}_b |$ and we fix an enumeration of this set as $\mathcal{E}_b =\{ e_0,\dots,e_{N_b-1} \}$. This enables us to define an enumeration over the full state-space $\mathcal{E} = \mathcal{E}_b \times \N^{d_f}_0$ by 
\begin{align}
\label{defn_mainenumeration}
\Psi(e,x) = N_b \Phi_{d_f}(x) + j, 
\end{align}
where $e = e_j \in \mathcal{E}_b$ and $x \in \N^{d_f}_0$. One can easily see that this map is a bijection between $\mathcal{E}_b \times \N^{d_f}_0$ and $\N_0$, and its inverse is given by
\begin{align}
\label{defn_mainenumeration_inv}
\Psi^{-1}(z) =(e_j,\Phi^{-1}_{d_f} (q) ), 
\end{align}
where $j$ is the remainder in the division of $z$ by $N_b$ and $q$ is the corresponding quotient. For the state-space $\mathcal{E}_b \times \N^{d_f}_0$ we define the trapezoidal truncation as
\begin{align}
\label{defn_section_full}
\mathcal{T}(C_l,C_r) =\mathcal{E}_b \times \{ x  \in \N^{d_f}_0 :   \Phi_{d_f}(C_l, {\bf 0} )  \leq \Phi_{d_f}(x) \leq  \Phi_{d_f}( {\bf 0},C_r) \},
\end{align}
where $C_l$ and $C_r$ are non-negative integers satisfying $C_l \leq C_r$ as before.

We now come to the definitions of finite state-space truncations $\mathcal{E}_i$-s and their enumerating functions $\phi_i$-s. Let $\{ C_{l,i} : i=1,2,\dots\}$ and $\{ C_{r,i} : i=1,2,\dots\}$ be monotonic sequences of non-negative integers that satisfy $C_{l,i} \leq C_{r,i}$ for each $i$ along with the limits
\begin{align}
\label{cutofflimits}
\lim_{i \to \infty} C_{l,i} = 0 \qquad \textnormal{and} \qquad  \lim_{i \to \infty} C_{r,i} = \infty.
\end{align}
For each $i=1,2,\dots$ we define the finite state-space truncation $\mathcal{E}_i$ as $\mathcal{T}(C_{l,i},C_{r,i}) $. Note that monotonicity of the left and right cut-off sequences  $\{ C_{l,i} \}$ and $\{ C_{r,i} \}$ along with \eqref{cutofflimits} ensures that $\{ \mathcal{E}_i : i =1,2,\dots \}$ is an increasing sequence of finite sets  that covers the full state-space $\mathcal{E}$ in the limit $i \to \infty$, as demanded by the sFSP Algorithm \ref{algo_sfsp}. Assuming that the Foster-Lyapunov function $V$ has the linear form \eqref{linearlyapunov}, we can choose the sequence $\{ \beta_i : i =1,2,\dots \}$ (see Remark \ref{thm:remark2}) in step 2 of Algorithm \ref{algo_sfsp} as $\beta_i =  C_{r,i}$.

In the case where the full state-space $\mathcal{E}$ is the non-negative integer orthant $\N^d_0$, the size of the truncated state-space $\mathcal{E}_i$ is
\begin{align}
\label{sizetruncatedstatepsace}
n_i  = | \mathcal{E}_i | =\Phi_d( {\bf 0}, C_{r,i}) -  \Phi_d(  C_{l,i}, {\bf 0}) + 1
\end{align}
and we enumerate the set $\mathcal{E}_i$ using the map $\phi_i : \mathcal{E}_i \to \{0,1,\dots, n_i-1\}$ given by
\begin{align*}
\phi_i(x) = \Phi_d( x ) -  \Phi_d(  C_{l,i}, {\bf 0}).
\end{align*}
Based on this enumeration the transition rate matrix $\bar{Q}_i $ for the projected CTMC over the truncated state-space $\mathcal{E}_i$ (see step 5 in Algorithm \ref{algo_sfsp}) can be constructed with Algorithm \ref{algoratematrixcreate}. In the other situation where the irreducible state-space $\mathcal{E}$ has the form \eqref{irredstatespaceform} for some finite non-empty set $\mathcal{E}_b = \{ e_0,\dots,e_{N_b-1} \}$ with $N_b =| \mathcal{E}_b |$ elements, the size of the truncated state-space $\mathcal{E}_i$ is
\begin{align*}
n_i  = | \mathcal{E}_i | = N_b \left( \Phi_{d_f}( {\bf 0}, C_{r,i}) -  \Phi_{d_f}(  C_{l,i}, {\bf 0}) + 1 \right)
\end{align*}
and we enumerate the set $\mathcal{E}_i$ using the map $\phi_i : \mathcal{E}_i \to \{0,1,\dots, n_i-1\}$ given by
\begin{align*}
\phi_i(e,x) = \Psi( e, x ) -  \Psi( e_0,C_{l,i}, {\bf 0}),
\end{align*} 
where $\Psi$ is the map defined by \eqref{defn_mainenumeration}. The transition rate matrix $\bar{Q}_i $ for the projected CTMC over the truncated state-space $\mathcal{E}_i$ can be constructed using Algorithm \ref{algoratematrixcreate} with some minor changes.

\begin{algorithm}[H]  
\caption{ Constructs the transition rate matrix according to \eqref{defn:qnbar} for the projected CTMC for network $\mathcal{R}$, over the truncated state-space $\mathcal{E}_i$ with a designated state $x_\ell \in \mathcal{E}_i$ .}      
 \label{algoratematrixcreate}
 \begin{algorithmic}[1]
 \Ensure The set $\mathcal{E}_i$ is a trapezoidal truncation $\mathcal{T}(C_{l,i},C_{r,i})$ given by \eqref{defn_section} for non-negative integers $C_{l,i}$ and $C_{r,i}$.
\Function{CreateRateMatrix}{$ \mathcal{R}, \mathcal{E}_i,x_\ell$}   
\State Let $n_i =| \mathcal{E}_i|$ be the size of the truncated state-space $\mathcal{E}_i$ given by \eqref{sizetruncatedstatepsace}. 
\State Initialize $\bar{Q}$ to be the $n_i \times n_i$ matrix of all zeros.
\State Set $m_0 = \Phi_d(C_{l,i}, {\bf 0})$
\State Set $l = \Phi_d(x_\ell) - m_0$ to be the \emph{address} of the designated state.
\For{$m=0,1,\dots,(n_i-1)$}
\State Set $y_m = \Phi^{-1}_d(m+m_0)$, $\lambda_0(y_m) = \sum_{k=1}^K \lambda_k(y_m)$ and $\bar{Q}_{mm} = -\lambda_0(y_m)$.
\For{$k=1,\dots,K$}
\If{$\lambda_k(y_m) > 0$}
\State Set $z = (y_m +\zeta_k)$ and $j = \Phi_d(z) - m_0$.
\If{$j < n$}
\State Set $\bar{Q}_{mj} = \lambda_k(y_m)$
\Else
\State  Set $\bar{Q}_{m l } = \bar{Q}_{m l }  + \lambda_k(y_m)$.
\EndIf
\EndIf
\EndFor
\EndFor
\State \Return the transition rate matrix $\bar{Q}_i = \bar{Q}$.
\EndFunction
\end{algorithmic}
\end{algorithm}

\subsection{Implementation Details}\label{sec:implementation1}

We now provide some details on our computer implementation of sFSP Algorithms \ref{algo_sfsp} and \ref{algoratematrixcreate}, and discuss the related issues. Note that the size $n_i$ of the truncated state-space $\mathcal{E}_i$ can be very large, causing problems in storing the $n_i \times n_i$ transition rate matrix $\bar{Q}_i$, and also in solving the linear-algebraic system \eqref{defn:stationarydistribution} to obtain $\bar{\pi}_i$. Note however that out of $n^2_i$ entries in matrix $\bar{Q}_i$, at most $n_i (K+1)$ entries can be non-zero, where $K$ is the number of reactions which is typically much smaller than $n_i$. Hence $\bar{Q}_i$ is an \emph{extremely sparse} matrix and this sparsity can be exploited for storing matrix $\bar{Q}_i$ and for finding the vector $\bar{\pi}_i$.

 Another issue that commonly arises is that for states with large components, the propensity functions take very high values which causes the matrix $\bar{Q}_i$ to have very large entries. This creates numerical issues while solving the linear-algebraic system \eqref{defn:stationarydistribution} for computing $\bar{\pi}_i$. A simple way to circumvent this problem is to \emph{scale} the matrix $\bar{Q}_i$ by its diagonal entries and apply the same scaling to the solution of the linear-algebraic system to recover $\bar{\pi}_i$. In other words, matrix $\bar{Q}_i$ is constructed by modifying Algorithm \ref{algoratematrixcreate} by setting $Q_{mm}$ to $-1$ in step 7 and by replacing $\lambda_k(y_m)$ with $\lambda_k(y_m)/\lambda_0(y_m)$ in steps 12 and 14. Such a scaling is allowed because the state-space $\mathcal{E}$ is irreducible for the original CTMC and hence any $y_m \in \mathcal{E}$ cannot be an \emph{absorbing state} and so $\lambda_0(y_m) = \sum_{k=1}^K \lambda_k(y_m)$ is nonzero. While constructing matrix $\bar{Q}_i$ we must also store the values $\lambda_0(y_m)$ for $m=0,1,\dots,n_i$. These values help in recovering $\bar{\pi}_i$ from the solution $\hat{\pi}_i$ of the linear-algebraic system solved in step 6 of Algorithm \ref{algo_sfsp}
 \begin{align*}
\bar{\pi}_{im} =  \frac{\hat{\pi}_{im} }{\lambda_0(y_m)}.
\end{align*}
Of course $\bar{\pi}_{i}$ is then normalized (step 7 of Algorithm \ref{algo_sfsp}) to ensure that its component-sum is $1$ and it represents a valid stationary distribution.

In our setup we implement the main sFSP method (Algorithm \ref{algo_sfsp}) in Matlab but we delegate the construction of the transition rate matrix $\bar{Q}_i$ to a C++ program that implements Algorithm \ref{algoratematrixcreate}. Once constructed, this matrix is imported into the sFSP Matlab program as a \emph{sparse} matrix. The linear-algebraic system \eqref{defn:stationarydistribution} for this matrix is solved by computing the eigenvector corresponding to the smallest-magnitude eigenvalue (i.e.\ $0$) using the {\bf eigs} function in Matlab. This function performs an Arnoldi iterative procedure \cite{Arnoldi} to efficiently compute a subset of eigenvalues and eigenvectors for large sparse matrices. It also allows us to pass a starting vector for the Arnoldi procedure. In our implementation we use the stationary distribution vector $\bar{\pi}_{i-1}$ obtained in iteration $(i-1)$ as the starting vector in iteration $i$\footnote{In the first iteration $i=1$, the starting vector is chosen to correspond to the uniform stationary distribution over the first state-space truncation $\bar{ \mathcal{E} }_1$}. For the sFSP implementation considered in this section, we use the scaled version of matrix $\bar{Q}_i$ as described above.

\subsection{Computational Examples} \label{sec:examples}

In this section we illustrate our simple implementation of sFSP using examples from systems biology. In all the considered examples, sFSP is applicable because with results in \cite{GuptaIrred} and \cite{GuptaPLOS} we can verify that the theoretical conditions required by sFSP (see Theorem \ref{thm:mainresult}) are satisfied. Moreover for all the examples, we fix the acceptable threshold level $\epsilon$ (see step 4 of Algorithm \ref{algo_sfsp}) to be $10^{-10}$, and we specify the increasing family of trapezoidal state-space truncations $\{ \mathcal{E}_i = \mathcal{T}(C_{l,i},C_{r,i})  \}$ via a pair of monotonic cut-off sequences $\{ C_{l,i}\}$ and $\{ C_{r,i}\}$ that satisfy $C_{l,i} \leq C_{r,i}$ for each $i$ along with the limits \eqref{cutofflimits}. The choice of these sequences can have a big influence on the overall performance of sFSP and especially the number of iterations it needs to terminate. Recall that $C_{l,i}$ and $C_{r,i}$ can be interpreted as bounds on the component-sum of states in the trapezoidal truncation $\mathcal{E}_i$ (see Section \ref{sec:enum}). Therefore we can use \emph{crudely} estimated values of the mean and standard deviation of the state component-sum at stationary, as a guidance for selecting these cut-off sequences. These crude estimates can be obtained with a few sample trajectories of the original CTMC generated with Gillespie's SSA \cite{GP}.

Since the two main steps of sFSP, viz.\ constructing the rate matrix $\bar{Q}_i$ and solving the linear-algebraic for $\bar{\pi}_i$, are performed on two separate computing platforms (C++ and Matlab), we will report the CPU times\footnote{All the computations for this simple implementation of sFSP were performed on an Apple machine with 2.9 GHz Intel Core i5 processor.} for both these steps individually for each iteration $i$. The total CPU time needed for an iteration is approximately the sum of these two times, and we will plot it along with the convergence factor $\gamma_i$, as a function of the iteration counter $i$, to show how they change as the truncated state-space $\mathcal{E}_i$ expands in size. For the computation of convergence factors we choose $\beta_i =  C_{r,i}$ in step 2 of Algorithm \ref{algo_sfsp} (see Section \ref{sec:enum}).

\subsubsection{Gene-expression network} \label{sec:genex}
Our first example is the gene-expression network given in \cite{MO}, where molecules of the \emph{messenger} RNA or mRNA (denoted by $M$) are created by a gene, and these mRNA molecules catalytically produce molecules of some protein (denoted by $P$). Molecules of both these species can degrade spontaneously. This two-species network has the following four reactions:
\begin{align*}
\emptyset \stackrel{\theta_1}{\longrightarrow} M , \quad     M \stackrel{\theta_2}{\longrightarrow}M + P,    \quad  M \stackrel{\theta_3}{\longrightarrow}\emptyset  \quad    \textrm{and} \quad  P  \stackrel{\theta_4}{\longrightarrow} \emptyset.
\end{align*}
The propensity functions are given by mass-action kinetics \eqref{massactionkinetics} and $\theta_i$-s denote the associated rate constants. We assume that the values of these rate constants are given by $\theta_1= 50$, $\theta_2 = 4$, $\theta_3 = 0.5$ and $\theta_4 = 0.2$. 

For the CTMC model of this network, the state-space $\mathcal{E} = \N^2_0$ is irreducible and so it can be enumerated with the Cantor Pairing function $\Phi_2$ (see Section \ref{sec:enum}). We apply sFSP to this network to obtain an estimate of the stationary probability distribution $\pi$. The cut-off sequences $\{C_{l,i}\}$ and $\{ C_{r,i} \}$ that define the trapezoidal truncation $\mathcal{E}_i = \mathcal{T}(C_{l,i},C_{r,i})$ for iteration $i$ are chosen as
\begin{align*}
C_{l,i} = \max\{ \hat{\mu} - 2\hat{\sigma} i,0  \} \quad \textnormal{and} \quad C_{l,r} = \hat{\mu} + 2\hat{\sigma} i, 
\end{align*}
where $ \hat{\mu} =2100$ and $\hat{\sigma} = 120$, are crudely estimated values of the mean and standard deviation of the state component-sum at stationarity, and these are obtained with a few SSA-generated trajectories of the CTMC. The designated state we select for sFSP is $(0, \hat{\mu} )$, which corresponds to $0$ mRNA molecules and $\hat{\mu} =2100$ protein molecules.

The performance of sFSP on the gene-expression network is summarized in Table \ref{table_sfsp_ge}, where for each iteration $i$, the cut-off values ($C_{l,i}$ and $C_{r,i}$), the truncated state-space size ($n_i= | \mathcal{E}_i |$), the convergence factor $\gamma_i$ and the CPU times for the two main sFSP steps are provided. One can see that sFSP terminated in $5$ iterations and overall it required 615 seconds of CPU time. To assess the accuracy of sFSP, we also estimate $\pi$ using $10^6$ CTMC trajectories simulated with SSA in the time-interval $[0,100]$. This SSA-based estimation was implemented in C++ and it needed 7246 seconds of CPU time which is much higher than the 615 seconds needed for sFSP. 

Note that the size of the truncated state-space $n_i$ is increasing linearly with $i$ and hence the size of the $n_i \times n_i$ rate matrix $\bar{Q}_i$ is increasing quadratically with $i$. So we would expect the CPU time for constructing $\bar{Q}_i$ and solving the linear-algebraic system for $\bar{\pi}_i$, to also increase quadratically with $i$. However this is not the case and the two CPU times only increase linearly (see Table \ref{table_sfsp_ge}). This is because matrix $\bar{Q}_i$ is extremely sparse with only $n_i K$ non-zero entries, where $K=4$ is the number of reactions. This sparsity is exploited in our implementation of sFSP for both constructing the matrix and solving the linear-algebraic system.

\begin{table}[h]
\begin{center}
\begin{tabular}{|c|c|c|c|c|c|c|}
\hline
Iteration & \multicolumn{2}{|c|}{Cut-offs}  & State-space size & Convergence factor &  \multicolumn{2}{|c|}{CPU Time (seconds) }  \\
\hline
$i$ & $C_{l,i}$ & $C_{r,i}$ & $n_i$& $ \gamma_i =  r^{(i)}_{ \textnormal{out} } C_{r,i} $ & Constructing $\bar{Q}_i$  &Finding $\bar{\pi}_i$   \\
\hline
\hline
$1$ & $1860$ & $2340$ & $1,008,240$& $2.541 \times 10^3$ &  13.6  & 18.7   \\
$2$ & $1620$ & $2580$ & $2,016,480$& $ 0.278$ &  27.5  & 46.4   \\
$3$ & $1380$ & $2820$ & $3,024,720$ & $7.473 \times 10^{-5}$ &  40.4  & 75.6   \\
$4$ & $1140$ & $3060$ & $4,032,960$ & $2.292 \times 10^{-9}$ &  53.6  & 110.7   \\
$5$ & $900$ & $3300$ & $5,041,200$ & $7.336 \times 10^{-15}$ &  68.3 & 159.6 \\
\hline
\end{tabular}
\end{center}
\caption{Application of sFSP on the gene-expression network. The transition rate matrix $\bar{Q}_i$ is constructed in C++ while its stationary distribution is found in Matlab.}
\label{table_sfsp_ge}
\end{table}

\begin{figure}[ht!]
\centering
\frame{\includegraphics[width=0.98\textwidth]{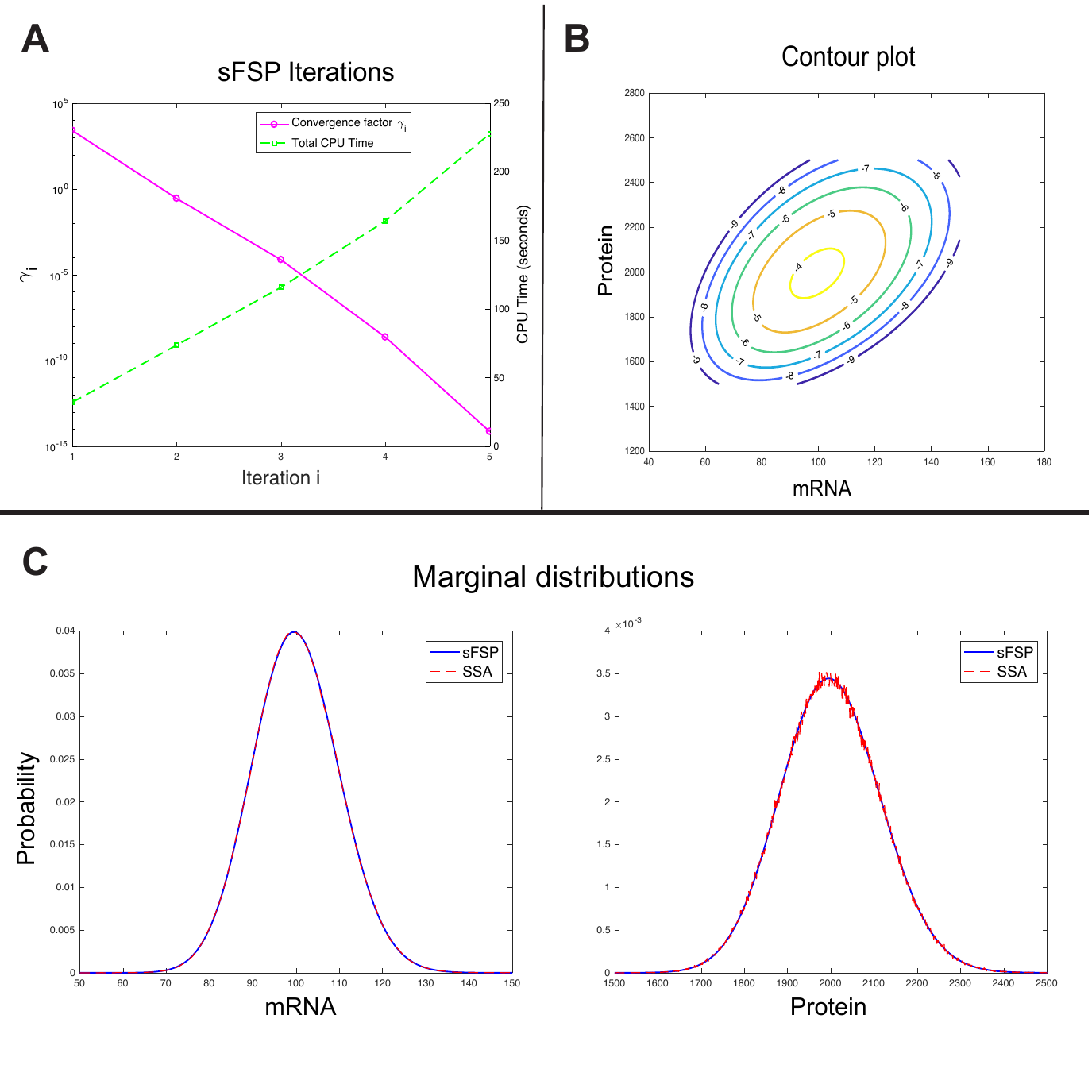}}
\caption{Application of sFSP on the gene-expression network. Panel {\bf A} plots the convergence factor $\gamma_i$ (in log-scale) and the total CPU time (calibrated against the right $y$-axis) as a function of the iteration counter $i$. This CPU time includes the time required for both rate matrix construction (performed with C++) and solving for the stationary distribution (performed with {\bf eigs} function in Matlab). Panel {\bf B} displays the contour plot for the estimated joint stationary distribution of mRNA and protein copy-numbers. This plot is in \emph{log-scale} which means that the contour level $-x$ corresponds to the probability of $10^{-x}$. Note that this contour plot indicates that the joint distribution is unimodal. In panel {\bf C} the estimated marginal stationary distribution for both mRNA and protein copy-numbers are plotted and also compared with the distributions estimated with Gillespie's SSA.}
\label{figure:sfsp_genex} 
\end{figure}

The linear increase in the required CPU time can be seen from Figure \ref{figure:sfsp_genex}{\bf A}. Here the convergence factor $\gamma_i$ is also plotted in log-scale and the \emph{almost} linear decay shows that the convergence factor drops exponentially to zero as the truncated state-space expands iteratively. Such an exponential decay is perhaps due to the fact that the joint stationary distribution is unimodal, as indicated by the contour plot in Figure \ref{figure:sfsp_genex}{\bf B}. This unimodality is also visible from the marginal distribution plots in Figure \ref{figure:sfsp_genex}{\bf C}. These sFSP-estimated marginal distribution plots are compared with the SSA-estimated distributions in  Figure \ref{figure:sfsp_genex}{\bf C} and they show a close match.     

\subsubsection{Toggle-Switch network} \label{sec:ts}

We now consider the example of the genetic toggle-switch network proposed by Gardner et.\ al.\ \cite{Gardner}. This network has two species ${\bf X}_1$ and ${\bf X}_2$ that are competing by \emph{repressing} each other's production. This repression is modeled through propensities given by nonlinear Hill functions \cite{Keener}. The network has four simple reactions    
\begin{align*}
\emptyset \stackrel{\lambda_1 }{\longrightarrow} {\bf X}_1 , \  \  {\bf X}_1 \stackrel{ \lambda_2 }{\longrightarrow} \emptyset,  \  \ \emptyset \stackrel{\lambda_3 }{\longrightarrow} {\bf X}_2  \   \textrm{ and }{\bf X}_2 \stackrel{ \lambda_4}{\longrightarrow} \emptyset,
\end{align*}
where the propensity functions $\lambda_i$-s are given by
\begin{align*}
\lambda_1(x_1,x_2) = \frac{\alpha_1}{ 1 +x_2^{\beta} }, \ \ \lambda_2(x_1,x_2) = \alpha_2 x_1 , \ \  \lambda_3(x_1,x_2) = \frac{\alpha_3}{ 1 +x_1^{\gamma} } \   \textrm{ and } \ \ \lambda_4(x_1,x_2) = \alpha_ 4 x_2. 
\end{align*}
Here $x_1$ and $x_2$ denote the copy-numbers of ${\bf X_1}$ and ${\bf X_2}$ respectively. For our computations we set $\alpha_1 =500 $, $\alpha_2 = 0.3$, $\alpha_3 = 200$, $\alpha_4 = 0.4$, $\beta = 1.5$ and $\gamma = 1$.

For the CTMC model of this network, the state-space $\mathcal{E} = \N^2_0$ is irreducible, and we apply sFSP with trapezoidal truncations using the cut-off sequences $\{C_{l,i}\}$ and $\{ C_{r,i} \}$ 
\begin{align*}
C_{l,i} = \max\{ \hat{\mu} - 0.5 \hat{\sigma} i,0  \} \quad \textnormal{and} \quad C_{l,r} = \hat{\mu} + 0.5 \hat{\sigma} i, 
\end{align*}
at iteration $i$, where $ \hat{\mu} =1110$ and $\hat{\sigma} = 500$, are crude SSA-based estimates of the mean and standard deviation of the state component-sum at stationarity. The designated state we select for sFSP is $(0, \hat{\mu} )$, which corresponds to $0$ molecules of ${\bf X_1}$  and $\hat{\mu} =1110$ molecules of ${\bf X_2}$.

The performance of sFSP on the Toggle-Switch network is summarized in Table \ref{table_sfsp_ts}, where for each iteration $i$, the cut-off values, the truncated state-space size, the convergence factor and the CPU times for the two main sFSP steps are provided. In this example, sFSP terminated in $5$ iterations and overall it required 322 seconds of CPU time. In comparison, the SSA-based estimation of $\pi$, implemented in C++, using $6 \times 10^6$ CTMC trajectories simulated in the time-interval $[0,100]$, needed 42192 seconds of CPU time. 

\begin{table}[ht!]
\begin{center}
\begin{tabular}{|c|c|c|c|c|c|c|}
\hline
Iteration & \multicolumn{2}{|c|}{Cut-offs}  & State-space size & Convergence factor &  \multicolumn{2}{|c|}{CPU Time (seconds) }  \\
\hline
$i$ & $C_{l,i}$ & $C_{r,i}$ & $n_i$& $ \gamma_i = r^{(i)}_{ \textnormal{out} } C_{r,i} $ & Constructing $\bar{Q}_i$  &Finding $\bar{\pi}_i$   \\
\hline
\hline
$1$ & $860$ & $1360$ & $555,250$& $ 1.03 \times 10^{3}  $ & 7.6 & 13.1  \\
$2$ & $610$ & $1610$ & $1,110,500$& $3.87  \times 10^{2}$ &  14.8  & 24.9  \\
$3$ & $360$ & $1860$ & $1,665,750$& $4.14 \times 10^{1}$ &  23.3  & 37.2   \\
$4$ & $110$ & $2110$ & $2,221,000 $& $1.37  \times 10^{-1}$ & 30.63  & 55.9  \\
$5$ & $0$ & $2360$ & $2,785,980$& $1.51 \times 10^{-53} $ &  37.66  & 77.1  \\
\hline
\end{tabular}
\end{center}
\caption{Application of sFSP on the Toggle-Switch network. The transition rate matrix $\bar{Q}_i$ is constructed in C++ while its stationary distribution is found in Matlab.}
\label{table_sfsp_ts}
\end{table}

\begin{figure}[h!]
\centering
\frame{\includegraphics[width=0.98\textwidth]{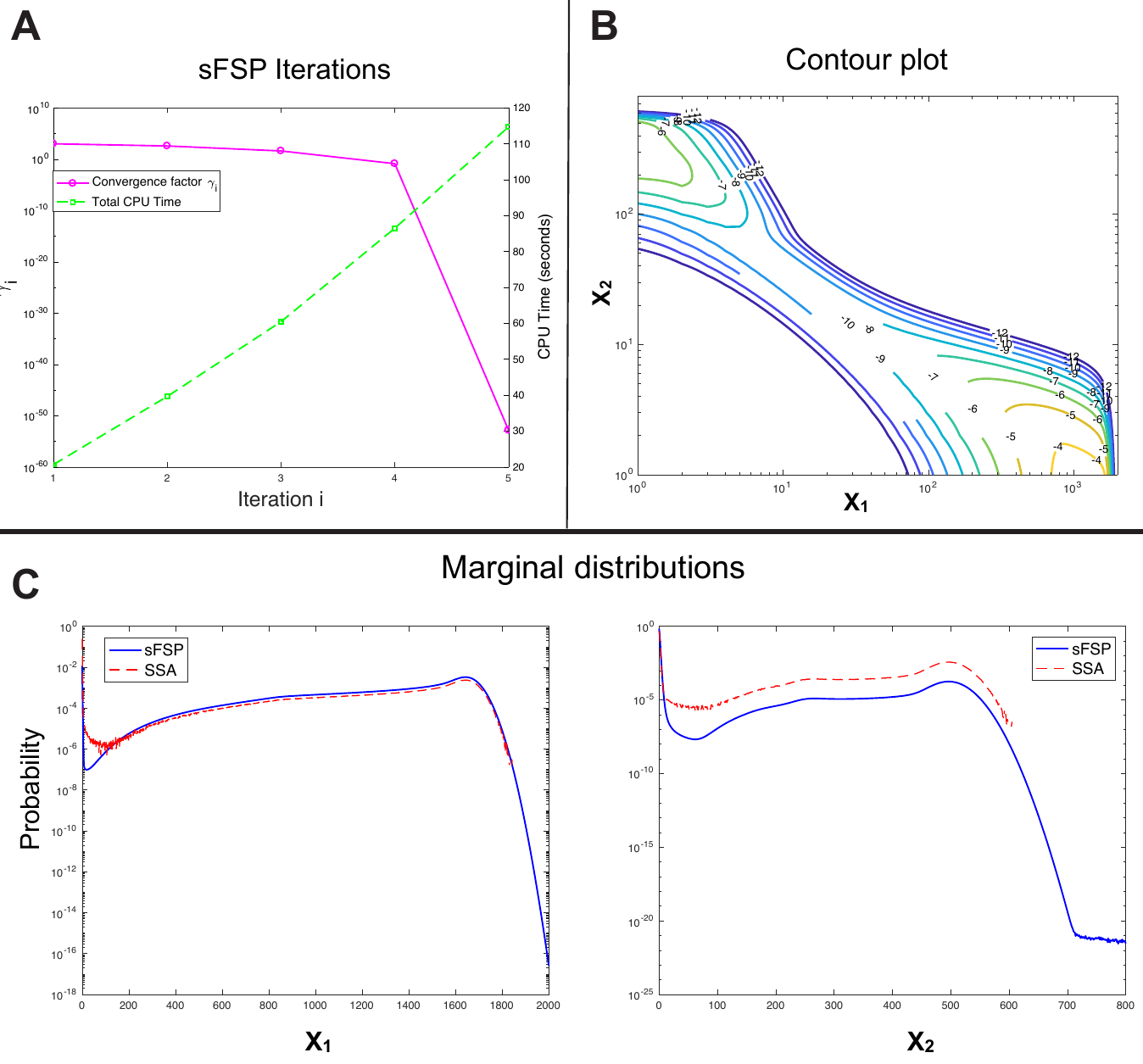}}
\caption{Application of sFSP on the Toggle-Switch network. Panel {\bf A} plots the convergence factor $\gamma_i$ (in log-scale)  and the total CPU time (calibrated against the right $y$-axis) as a function of the iteration counter $i$. This CPU time includes the time required for both rate matrix construction (performed with C++) and solving for the stationary distribution (performed with {\bf eigs} function in Matlab). Panel {\bf B} displays the contour plot for the estimated joint stationary distribution of the copy-numbers of the two species. This plot is in \emph{log-scale} which means that the contour level $-x$ corresponds to the probability of $10^{-x}$. Note that this contour plot indicates that the joint distribution is bimodal. In panel {\bf C} the estimated marginal stationary distribution for both the species copy-numbers are plotted and also compared with the distributions estimated with Gillespie's SSA.}
\label{figure:sfsp_ts} 
\end{figure}

As in the previous example, the required CPU time increases almost linearly with iteration $i$ and the convergence factor $\gamma_i$ decreases slowly for the first four iterations and then plummets to nearly $0$ in the fifth iteration (see Figure \ref{figure:sfsp_ts}{\bf A}). The contour plot for the joint stationary distribution estimated by sFSP is shown in Figure \ref{figure:sfsp_ts}{\bf B} and it indicates that this distribution is \emph{bimodal} with each mode corresponding to one of the species being dominant. In Figure \ref{figure:sfsp_ts}{\bf C} we plot the sFSP-estimated marginal stationary distributions for the copy-numbers of the two species and compare them with the SSA-estimated marginal stationary distributions. One can clearly see that unlike sFSP, SSA fails to adequately capture the stationary distribution in the low-probability regions of the state-space even though a large sample of size $6$ million is used. These statistical errors and other numerical issues associated with computing very low probabilities, may explain the slight discrepancy in the sFSP and SSA estimated marginal distribution for species ${\bf X_2}$ (see Figure \ref{figure:sfsp_ts}{\bf C}).

\subsubsection{Pap-Switch network} \label{sec:ps}

\begin{figure}[ht!]
\centering
\frame{\includegraphics[width=0.98\textwidth]{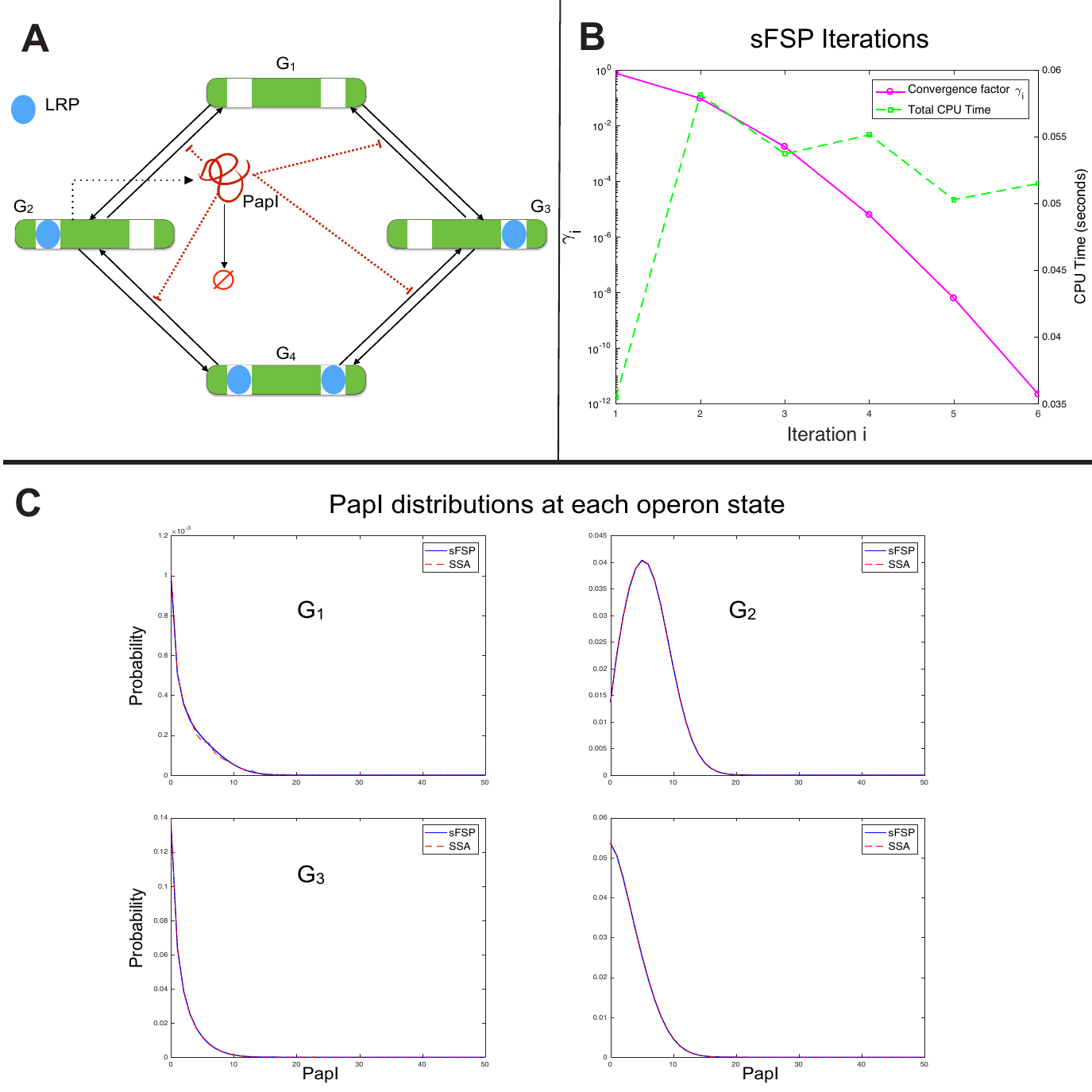}}
\caption{Panel {\bf A} depicts the Pap-Switch network with a \emph{pap} operon switching between four states $G_1,\dots,G_4$ and producing the PapI protein in state $G_2$. This protein represses certain operon-state transitions as shown by the \emph{dotted red lines}. sFSP is applied to this network and panel {\bf B} plots the convergence factor $\gamma_i$ (in log-scale) and the total CPU time (calibrated against the right $y$-axis) as a function of the iteration counter $i$. The estimated stationary distributions for PapI copy-numbers at each operon state are plotted in panel {\bf C} and also compared with the distributions estimated with Gillespie's SSA.}
\label{figure:sfsp_ps} 
\end{figure}

We now consider the Pap epigenetic switch whose finite-time CME was solved in \cite{FSP} with the FSP method. This stochastic switch is responsible for deciding whether or not \emph{E. coli} will develop hairlike structures called pili. The Pap-switch network is illustrated in Figure \ref{figure:sfsp_ps}{\bf A} and it consists of a single \emph{pap} operon $G$ that can exist in four states $G_1,G_2,G_3$ and $G_4$ determined by the binding sites occupied by the leucine-responsive regulatory protein (LRP) molecules. When the operon is in state $G_2$, it can produce a local regulatory protein called \emph{PapI} which represses the unbinding of the LRP molecules from the operon binding sites. This $PapI$ protein is allowed to degrade spontaneously at a certain rate. As in \cite{FSP} we assume that the number of LRP molecules is fixed at $100$. The dynamics of the copy-numbers of the five species $G_1,G_2,G_3,G_4$ and $PapI$ in the Pap-Switch network can be modeled with $10$ reactions described in Table \ref{tab:pap}.

\begin{table}[h!]
\begin{align*}
\begin{array}{|c|l|l|}\hline 
\textnormal{No.} & \textnormal{Reaction} &  \textnormal{Propensity} \\ \hline
1 & G_1 +\textnormal{[LRP]}  \longrightarrow  G_2  &  \lambda_1(x) = \textnormal{[LRP]} x_1 \\ 
2 & G_2  \longrightarrow  G_1 + \textnormal{[LRP]} &  \lambda_2 (x) = \left(0.25 + 2.25/(1+x_5)\right)x_2 \\
 3 &G_1 +\textnormal{[LRP]}  \longrightarrow  G_3 &  \lambda_3 (x) = \textnormal{[LRP]} x_1 \\
 4 &G_3  \longrightarrow  G_1 + \textnormal{[LRP]} &  \lambda_4 (x) = \left(1 + 0.2/(1+x_5)\right)x_3 \\
 5 &G_2 +\textnormal{[LRP]}  \longrightarrow  G_4 &  \lambda_5 (x) = 0.01(\textnormal{[LRP]}-1) x_2\\
 6 &G_4  \longrightarrow  G_2 + \textnormal{[LRP]} &  \lambda_6(x) = \left(1 + 0.2/(1+x_5)\right)x_4  \\
 7 &G_3 +\textnormal{[LRP]}  \longrightarrow  G_4 &  \lambda_7 (x) = 0.01(\textnormal{[LRP]}-1) x_2\\
 8 &G_4  \longrightarrow  G_3 + \textnormal{[LRP]}t&  \lambda_8 (x) = \left(0.25 + 2.25/(1+x_5)\right)x_4 \\
 9 & G_2  \longrightarrow  G_2 + PapI  &  \lambda_9 (x) = 10 x_2 \\
 10 &  PapI \longrightarrow  \emptyset&  \lambda_{10} (x) = x_4 \\
\hline
\end{array}
\end{align*}
\caption{Reactions for the Pap-Switch. Here $\textnormal{[LRP]}  =100$ denotes the total number of LRP molecues and $x = (x_1,\dots,x_5)$ denotes the copy-numbers of the five species ordered as $G_1,G_2,G_3,G_4$ and $PapI$. Propensities of reactions $2,4,6$ and $8$ contain a term for the repression of LRP unbinding by $PapI$ molecules. 
}
\label{tab:pap}
\end{table}

For the CTMC model of this network, the state-space $\mathcal{E} = \mathcal{E}_b \times \N_0$ is irreducible, where $$\mathcal{E}_b = \{  (1,0,0,0), (0,1,0,0), (0,0,1,0), (0,0,0,1)\}$$ is the finite set which contains the dynamics of the copy-numbers $(x_1,x_2,x_3,x_4)$ of the four operon states $G_1,G_2,G_3$ and $G_4$. The copy-numbers of $PapI$ can take values in the whole set of non-negative integers $\N_0$. The state-space of the form $\mathcal{E} = \mathcal{E}_b \times \N_0$ can be enumerated using the function $\Psi$ (see \eqref{defn_mainenumeration}) with $N_b= 4$. Similarly the trapezoidal truncations $\mathcal{E}_i$-s can be defined as \eqref{defn_section_full}. In our application of sFSP for this network we construct these truncations using the cut-off sequences $\{C_{l,i}\}$ and $\{ C_{r,i} \}$ specified by  
\begin{align*}
C_{l,i} = 0 \quad \textnormal{and} \quad C_{l,r} = \hat{\mu} + \hat{\sigma} i, 
\end{align*}
at iteration $i$, where $ \hat{\mu} =4$ and $\hat{\sigma} = 3$, are coarse SSA-based approximations of the mean and standard deviation of the $PapI$ copy-numbers. Note that due to the low copy-numbers involved we fix the left cut-off point $C_{l,i} $ to be zero for all the iterations. Also the designated state we select for sFSP is $(1,0,0,0,0)$, which corresponds to the operon being in state $G_1$ and $PapI$ having $0$ molecules.

The performance of sFSP on the Pap-Switch network is summarized in Table \ref{table_sfsp_paps}, where for each iteration $i$, the cut-off values, the truncated state-space size, the convergence factor and the CPU times for the two main sFSP steps are provided. For this network, sFSP took $6$ iterations to terminate and overall it required only $0.304$ seconds of CPU time. By contrast, the SSA-based estimation of the stationary distribution, implemented in C++, with $10^6$ CTMC trajectories generated in the time-period $[0,100]$, required 134 seconds of CPU time.

\begin{table}[h]
\begin{center}
\begin{tabular}{|c|c|c|c|c|c|c|}
\hline
Iteration & \multicolumn{2}{|c|}{Cut-offs}  & State-space size & Convergence factor &  \multicolumn{2}{|c|}{CPU Time (seconds) }  \\
\hline
$i$ & $C_{l,i}$ & $C_{r,i}$ & $n_i$& $ \gamma_i = r^{(i)}_{ \textnormal{out} } C_{r,i}  $ & Constructing $\bar{Q}_i$  &Finding $\bar{\pi}_i$   \\
\hline
\hline
$1$ & $0$ & $10$ & $44$& $ 7.72 \times 10^{-1} $   & 0.00062 & 0.0344   \\
$2$ & $0$ & $16$ & $68$& $ 9.64 \times 10^{-2}$ &  0.00117& 0.0566 \\
$3$ & $0$ & $22$ & $92$& $1.75 \times 10^{-2} $ & 0.001719 & 0.0517   \\
$4$ & $0$ & $28$ & $116$& $6.26  \times 10^{-6} $ &  0.002906&0.0519 \\
$5$ & $0$ & $34$ & $140$& $ 6.31 \times 10^{-9} $ &  0.003498& 0.0464 \\
$6$ & $0$ & $40$ & $164$& $2.23 \times 10^{-12}$ & 0.00255&0.0486 \\
\hline
\end{tabular}
\end{center}
\caption{Application of sFSP on the Pap-Switch network. The transition rate matrix $\bar{Q}_i$ is constructed in C++ while its stationary distribution is found in Matlab.}
\label{table_sfsp_paps}
\end{table}

In this example, the sizes of the truncated state-spaces are very small and so sFSP executes very quickly, causing the CPU times to vary non-monotonically with iteration $i$ while the convergence factor decreases almost exponentially (see Figure \ref{figure:sfsp_ps}{\bf B}). In Figure \ref{figure:sfsp_ps}{\bf C} we plot the sFSP-estimated stationary distributions for $PapI$ copy-numbers at each operon state $G_1,G_2,G_3$ and $G_4$. These are compared with the corresponding SSA-estimated stationary distributions and it can be seen from Figure \ref{figure:sfsp_ps}{\bf C} that the match is almost perfect.

\subsubsection{Self-activated gene expression} \label{sec:munsky}
We end this section with a simple but instructive example borrowed from \cite{fox2016finite}. Consider a gene whose protein output ${\bf X}$ can activate its own expression through a nonlinear feedback loop. A simple reaction network model for this would be
\begin{align*}
\emptyset \stackrel{ \lambda_1(x) }{ \longrightarrow } {\bf X}  \stackrel{ \lambda_2(x)}{ \longrightarrow }  \emptyset,
\end{align*}
where the propensity function for the degradation reaction is linear $\lambda_2(x) = \gamma x$ while the propensity function for the production reaction is given by a Hill-type function
\begin{align*}
\lambda_1(x) = k_1 + k_2 \left(  \frac{x^\alpha}{m^\alpha + x^\alpha}  \right).
\end{align*}
Here $x$ denotes the copy-number of protein ${\bf X}$. For our computations we set $k_1 = 20$, $k_2  = 125$, $\alpha = 5$, $m = 70$ and $\gamma = 1$.

\begin{figure}[ht!]
\centering
\includegraphics[width=1\textwidth]{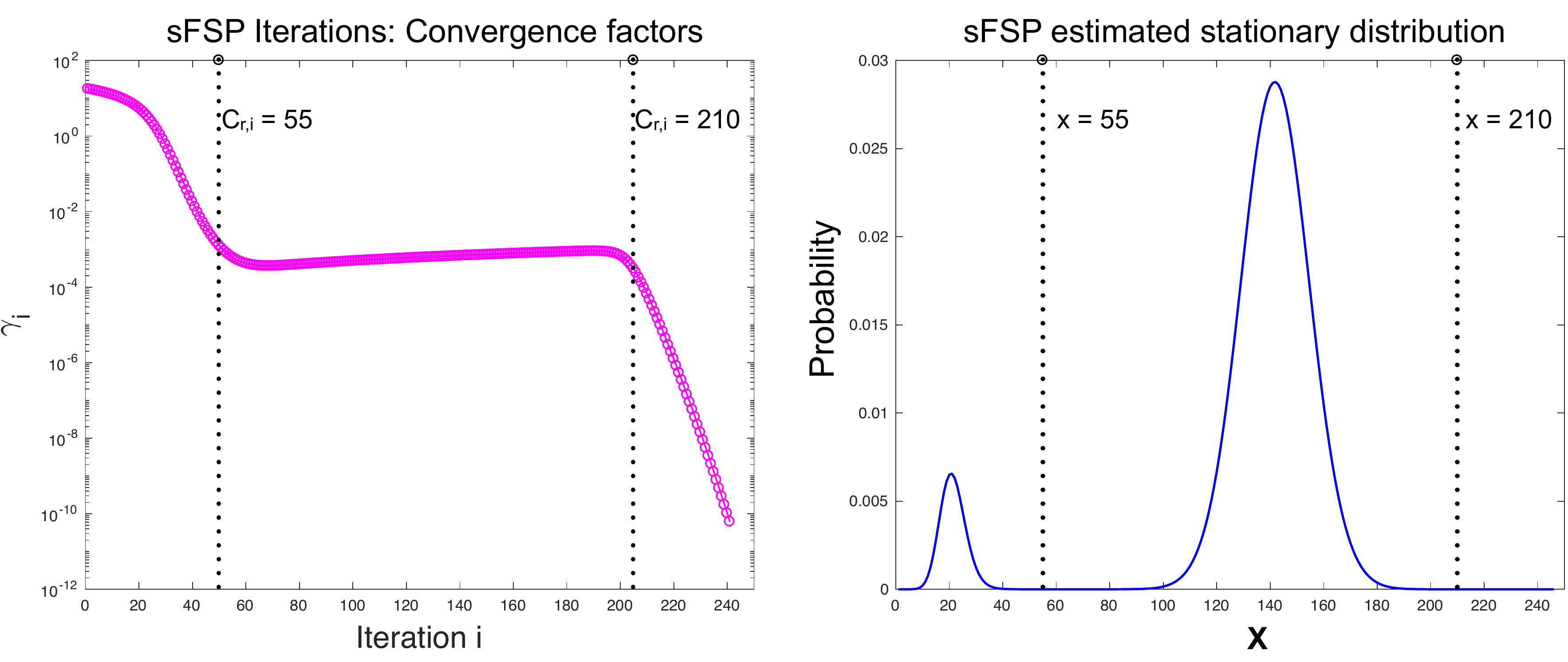}
\caption{Results from the application of sFSP are shown for the self-activated gene expression example in Section \ref{sec:munsky}. Here the truncated state-space at iteration $i$ is $\mathcal{E}_i = \{ 0,1,\dots,C_{r,i}\}$ with $C_{r,i} = (5+i)$. The stationary distribution is bimodal with most of the probability-mass concentrated in regions $R_1 = \{0,\dots, 55\}$ and $R_2 = \{80,\dots, 210\}$. The end-points of these two regions correspond to \emph{inflection} points for the relationship between the iteration counter $i$ and the convergence factor $\gamma_i$. In particular, the convergence factor $\gamma_i$ decays exponentially (i.e. linearly in the log-scale used above for the left plot) until the end-point of $R_1$ is reached (at $i = 50$ or $C_{r,i} = 55$). It then \emph{increases} slowly until the end-point of $R_2$ is reached (at $i = 205$ or $C_{r,i} = 210$), and thereafter it resumes its exponential decay at an even faster rate than in region $R_1$. 
}
\label{figure:sfsp_munsky} 
\end{figure}

As there is only one species, the trapezoidal truncation $\mathcal{E}_i = \mathcal{T}( C_{l,i} , C_{r,i} )$ is simply the set $\mathcal{E}_i = \{ C_{l,i} , C_{l,i}+1,\dots,C_{r,i}\}$. We choose $C_{l,i} = 0$ and $C_{r,i} = (5+i)$ at iteration $i$, and apply sFSP on this example with designated state $0$. The results are shown in Figure \ref{figure:sfsp_munsky}. Note that the stationary distribution is bimodal, with a \emph{small} peak around $20$ and a \emph{larger} peak around $145$. Most of the stationary probabilities are concentrated in two disjoint regions $R_1 = \{0,\dots, 55\}$ and $R_2 = \{80,\dots, 210\}$ around the two peaks. Observe that the end-points $x_1 = 55$ and $x_2 = 210$ of these two regions are \emph{inflection} or \emph{turning} points for the behavior of the convergence factor $\gamma_i$ with increasing iteration counter $i$ or expanding truncated state-space $\mathcal{E}_i$. The convergence factor $\gamma_i$ decays exponentially before $x_1$ and after $x_2$, but in the intermediate region $I = \{x_1+1,\dots, x_2-1 \}$ it shows a gradual increase. Further computations reveal that for iterations corresponding to this intermediate region, the outflow rate $r^{ (i) }_{ \textnormal{out} }$ remains approximately constant, and so the convergence factor $\gamma_i$ increases slowly due to scaling by the cut-off value $C_{r,i}$. This relationship between \emph{bimodality} of the stationary distribution and non-monotonicity of the convergence factor $\gamma_i$ is very interesting and should be investigated in a greater detail elsewhere.

\section{sFSP Algorithm: QTT Implementation}  \label{secqttsfsp}

The second implementation is motivated by the recently developed \emph{Quantized Tensor-Train} (QTT) version of FSP \cite{Kazeev}, which works with QTT representations of the transition rate matrix and its stationary distribution vector. The use of such representations expands the range of applicability of sFSP and we demonstrate this by applying sFSP on a network which is much larger than the networks considered in Section \ref{sec:examples}.

\subsection{The CME in QTT form}

A \emph{tensor} is essentially a multi-dimensional generalization of a two-dimensional matrix or a one-dimensional vector. A $d$-dimensional tensor $T$ of size ${\bm n} = n_1 \times \dots \times n_d$, represents a structured collection of real numbers given by $$\{ T(i_1,\dots,i_d):  0 \leq i_k \leq (n_k - 1) \quad \textnormal{for} \quad k=1,\dots,d \}. $$ 
Each dimension of this tensor $T$ is also called its \emph{mode}, and $n_1, \ldots, n_d$ denote the mode sizes. The tensor $T$ can also be viewed as a real-valued function over the $d$-dimensional \emph{hyper-rectangle}
\begin{align}
\label{lattice_tensor_defn}
\mathcal{E}_{ \bm n } = \bigotimes_{k=1}^d \left\{0,1,\dots, n_k -1\right\}
\end{align}
which is a subset of the non-negative integer orthant $\N^d_0$. 

Tensors are particularly well suited to express the CME since the system already has a physical interpretation as tensors, where each species corresponds to one tensor mode and for any mode $k$, its size $n_k$ serves as the strict upper-bound for the allowable copy-numbers for species $\mathbf{X}_k$. As in FSP \cite{FSP}, consider a CME over the truncated state-space $\mathcal{E}_{ \bm n }$ (see \eqref{cme_fsp} for example). The probability distribution $p_{\bm n}(t)$ of the random state-vector at time $t$ can be represented as a $d$-dimensional tensor of size ${\bf n}$ and the matrix $Q^T_{ \bm n }$ that captures its rate of change can be represented as a $2d$-dimensional tensor of size ${\bf n} \times {\bf n}$.    

The \emph{tensor train} (TT) representation of a $d$-dimensional tensor $T$ with size ${\bm n} = n_1 \times \dots \times n_d$ is given by
\begin{align*}
T(i_1, \ldots, i_d) = \sum\limits_{\alpha_0 = 1}^{r_0} \dots \sum\limits_{\alpha_{d} = 1}^{r_{d}}U_1( \alpha_0, i_1, \alpha_1) U_2(\alpha_1, i_2, \alpha_2)\dots U_{d-1}(\alpha_{d-1}, j_{d-1}, \alpha_{d-1}) \cdot U_d(\alpha_{d-1}, i_d, \alpha_d),
\end{align*}
where $r_0=r_d =1$ and for each $j =1,\dots,d$, $U_j$ is a three-dimensional tensor with size $r_{j-1} \times n_j  \times r_j$. The tensors $U_1$ to $U_d$ are called the core tensors and $r_1, \ldots r_{d-1}$ are referred to as tensor ranks. The TT-representation can potentially provide a high compression of the tensor, especially if the ranks are low. Most basic matrix-vector operations (like matrix-vector product, dot product, outer product etc.) can be applied directly on the compressed TT format (for details see \cite{oseledets2011tensor}). The complexity of these basic operations as well as the storage cost can be bound by $ n_{ \textnormal{max} } r_{ \textnormal{max} }^2d$ where $n_{ \textnormal{max} } =\max\{ n_1, \ldots n_d \}$ and $r_{ \textnormal{max} } =\max\{ r_1, \ldots, r_{d-1}\}$. Any tensor can be decomposed into the TT format by the TT-SVD algorithm \cite{oseledets2011tensor}, which is based on the \emph{Singular Value Decomposition} (SVD) for matrices. The TT format can be extended to the \emph{quantized tensor train} (QTT) format which provides another layer of compression by dividing each mode of the tensor into several \emph{virtual} modes that are then further compressed using tensor trains (see \cite{oseledets2010approximation} and \cite{oseledets2009approximation}).

In \cite{Kazeev} the authors show how the matrix $Q^T_{ \bm n }$ for the CME \eqref{cme_fsp} over the truncated state-space $\mathcal{E}_{ \bm n }$ \eqref{lattice_tensor_defn} can be directly constructed in the QTT format and thereafter used for efficiently solving the FSP and obtaining the transient CME solution $p_{\bm n}(t)$. The main observation underlying the QTT construction of $Q^T_{ \bm n }$ is that one can think of this matrix in terms of the spatial shift operator $\bm{S}_{\zeta_k}$, shifting a probability density tensor $p$ by the stoichiometry vector $\zeta_k$ for reaction $k$, and a multiplication operator $\bm{M}_{\lambda_k}$, multiplying a probability density tensor $p$ by the propensity function $\lambda_k$ for reaction $k$, i.e. 
\begin{align*}
\bm{S}_{\zeta_k} p(x) = p(x -\zeta_k) \quad \textnormal{and} \quad \bm{M}_{\lambda_k}p(x) = \lambda_k(x) p(x)
\end{align*}
for any $x \in \mathcal{E}_{ \bm n }$. Using these operators along with the identity operator $\mathbb{I}$, the matrix $Q^T_{ \bm n }$ can be expressed as
\begin{align}
\label{cme_qtt_decomp}
Q^T_{ \bm n } = \sum_{k=1}^K \left( \bm{S}_{\zeta_k} - \mathbb{I} \right) \circ \bm{M}_{\lambda_k},
\end{align}
and this form can be exploited for efficiently constructing the QTT representation of $Q^T_{ \bm n }$. As explained in \cite{Kazeev}, for mass action kinetics, the operator $\bm{M}_\lambda$ can be constructed by taking the outer products of state-vectors in $\mathcal{E}_{ \bm n }$ and the appropriate vector of ones $\mathbf{1}$, while the operator $\bm{S}_{\zeta_k}$ can be constructed as a matrix of zeros with a shifted diagonal of ones.

\subsection{Implementation Details} \label{impl_detail2}

In our QTT implementation of sFSP, we use a similar expression as \eqref{cme_qtt_decomp} to construct the QTT representation of the transpose $\bar{Q}^T_{ \bm n }$ of the transition rate matrix $\bar{Q}_{ \bm n }$ (see \eqref{defn:qnbar}) for our projected CTMC over the truncated state-space $\mathcal{E}_{ \bm n }$, where all the outgoing transitions are redirected to the designated state ${\bf 0}$ of all zeros. Using the QTT representation of $\bar{Q}^T_{ \bm n }$, the corresponding linear-algebraic system \eqref{defn:stationarydistribution} is directly solved in QTT format to yield the stationary probability distribution $\bar{\pi}_{ \bm n }$ in QTT format.

For solving the linear-algebraic system, we use the \emph{inverse iteration} approach (see \cite{ipsen1997computing}) which is known to have very good convergence properties and work well with tensor algebra \cite{rakhuba2016calculating}. In this approach a linear system of the form $A x = {\bf 0}$, for a singular matrix $A$, is solved by iteratively solving the linear systems
\begin{align*}
A x_{j} = x_{j-1} \quad \textnormal{for} \quad j=1,2,\dots
\end{align*}
starting with some initial guess $x_0$. The solution $x_j$ is suitably normalized before commencing iteration $(j+1)$. Generally this procedure requires very few iterations (like 2 or 3) to converge, and this convergence can be judged by checking that the distance between subsequent solutions $\|x_j - x_{j-1}\|$ is below some threshold level $\delta$.

In our setup we implement the QTT version of sFSP method (Algorithm \ref{algo_sfsp}) in Matlab, using Version 2.2 of the \texttt{qtt-toolbox} developed by I. Oseledets,  S. Dolgov, V. Kazeev, O. Lebedeva, and T. Mach \cite{qtttoolbox}. In particular the linear systems that arise in the inverse iteration procedure are solved using the function {\bf dmrg\_solve3.m} from this toolbox. For each sFSP iteration $i$, the initial guess for the inverse iteration procedure is chosen based on the estimate obtained in iteration $(i-1)$, as mentioned in Section \ref{sec:implementation1}. For the computational example we consider next, we found that only two inverse iterations were always sufficient to yield a convergent solution of the linear-algebraic system \eqref{defn:stationarydistribution} for the threshold level $\delta = 10^{-4}$.

\subsection{A Computational Example} \label{sec:examples2}
  
\begin{figure}[ht!]
\centering
\frame{\includegraphics[width=0.98\textwidth]{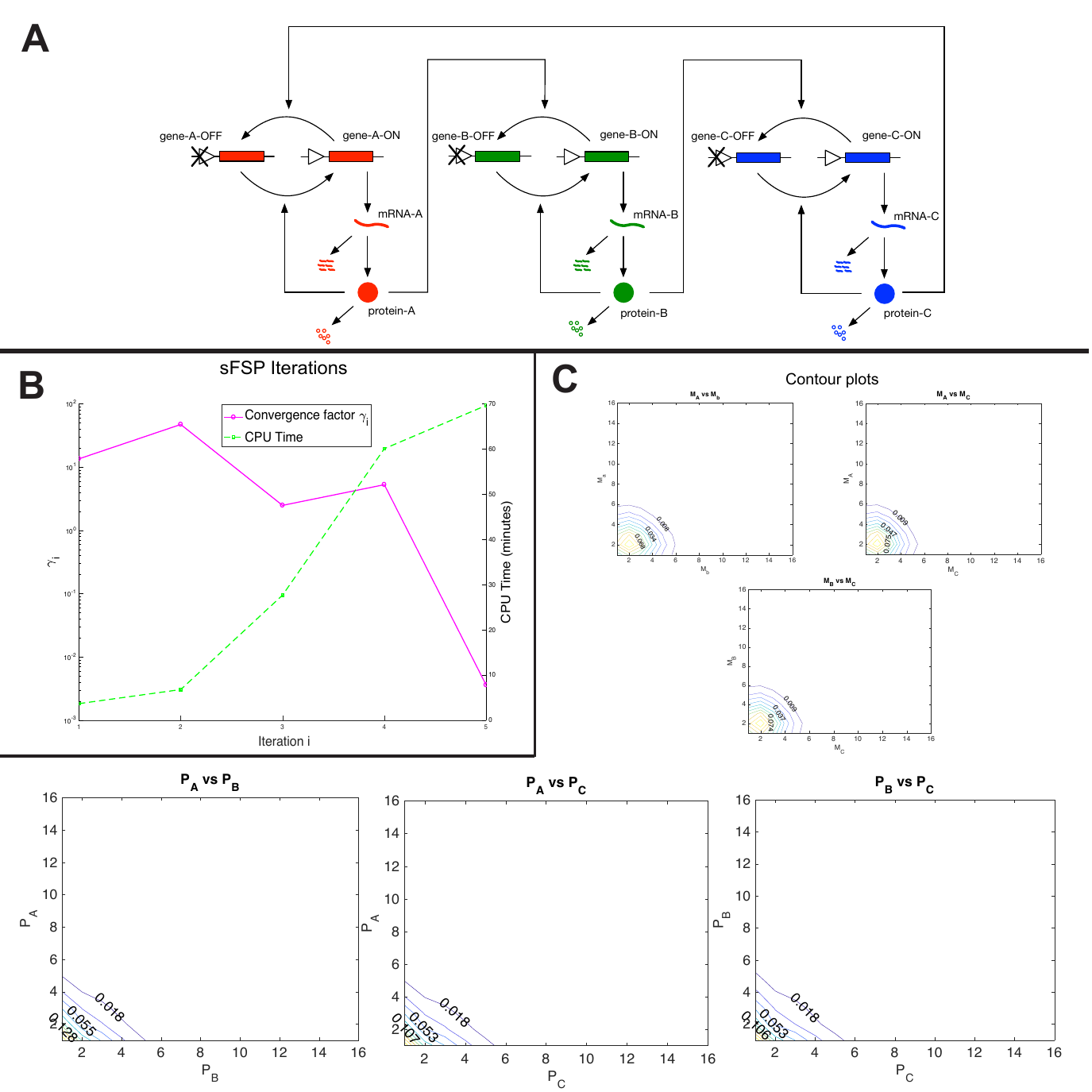}}
\caption{Panel {\bf A} depicts the triple-repressor model with three self-activating genes (A, B and C) that interact by repressing each other in a cyclic fashion via their corresponding proteins. We apply the QTT version of sFSP to this network and panel {\bf B} plots the convergence factor $\gamma_i$ (in log-scale) and the CPU time (calibrated against the right $y$-axis) as a function of the iteration counter $i$. Panel {\bf C} displays the contour plots for the estimated joint stationary distribution of the copy-numbers of various pairs of species. All the contour plots indicate unimodality of the joint distributions. 
}
\label{fig:repress1} 
\end{figure}

We now illustrate our QTT implementation on a toy example with features similar to the \emph{Repressilator} network given by Elowitz and Leibler \cite{elowitz2000synthetic}, which has three gene-expression modules (say A, B and C) that interact by mutual inhibition of each other in a cyclic fashion i.e.\ $A$ represses $B$, $B$ represses $C$ and $C$ represses $A$ (see Figure \ref{fig:repress1}{\bf A}). This inhibition is carried out by the corresponding proteins ($P_A$, $P_B$ and $P_C$) and it is achieved by enhancing the rate at which the inhibited gene becomes inactive (OFF) from an active (ON) state. Each protein also activates its own production by increasing the rate at which its gene switches ON from the OFF state. The mRNAs ($M_A$, $M_B$ and $M_C$) associated with the genes are only transcribed when the corresponding gene is in the ON state. Overall this network consists of $9$ species and $18$ reactions described in Table \ref{tab:repress1}. These $9$ species include the indicators for the three genes being in the ON state ($G_A^1$, $G_B^1$ and $G_C^1$), the three mRNAs ($M_A$, $M_B$ and $M_C$) and finally the three proteins ($P_A$, $P_B$ and $P_C$).

\begin{table}[h!]
\centering
\begin{tabular}{|c | l | l| }
\hline 
No. & Reaction & Propensity \\
\hline
1 & $G_A^0 \longrightarrow  G_A^1 $ &  $\lambda_{1}(x) = (10 + 1.5 x_7)(1-x_1)$ \\
2 & $G_A^1 \longrightarrow G_A^0 $ &  $\lambda_{2}(x) = (7 + 2 x_9)x_1$ \\
3 & $G_B^0 \longrightarrow  G_B^1$ &  $\lambda_{3}(x) = (9+ 4 x_8)(1-x_2)$ \\
4 & $G_B^1 \longrightarrow G_B^0$ &   $\lambda_{4}(x) = (10 + 4 x_7)x_2$ \\
5 & $G_C^0 \longrightarrow G_C^1$ &   $\lambda_{5}(x) = (11 + 1.5 x_9)(1-x_3)$ \\
6 & $G_C^1 \longrightarrow G_C^0 $ &   $\lambda_{6}(x) = (9 + 2 x_8)x_3$ \\
7 & $G_A^1 \longrightarrow G_A^1 + M_A$ & $\lambda_{7}(x) =  1.5 x_1$ \\
8 & $G_B^1 \longrightarrow G_B^1 + M_B$ & $\lambda_{8}(x) = 1x_2$ \\
9 & $G_C^1 \longrightarrow G_C^1 + M_C$ & $\lambda_{9}(x) = 1.1x_3$ \\
10 & $M_A \longrightarrow \emptyset $ &   $\lambda_{10}(x) = 0.5 x_4$ \\
11 & $M_B \longrightarrow \emptyset $ &   $\lambda_{11}(x) = 0.3 x_5$ \\
12 & $M_C \longrightarrow \emptyset $ &  $\lambda_{12}(x) = 0.425 x_6$  \\
13 & $M_A \longrightarrow M_A + P_A $ &   $\lambda_{13}(x) = 9.5 x_4$ \\
14 & $M_B \longrightarrow M_B + P_B $ &   $\lambda_{14}(x) = 11 x_5$ \\
15 & $M_C \longrightarrow M_C + P_C $ &   $\lambda_{15}(x) = 10 x_6$ \\
16 & $P_A \longrightarrow \emptyset $ &   $\lambda_{16}(x) = 14.5 x_7$ \\
17 & $P_B \longrightarrow \emptyset $ &   $\lambda_{17}(x) = 15 x_8$ \\
18 & $P_C \longrightarrow \emptyset $ &   $\lambda_{18}(x) = 11 x_9$ \\
\hline
\end{tabular}
\caption{Reactions for the triple-repressor model. Here $x = (x_1 , . . . , x_9 )$ denotes the copy-numbers of the 9 network species ordered as $G^1_A$, $G^1_B$, $G^1_C$, $M_A$, $M_B $, $M_C$, $P_A$, $P_B$ and $P_C$. Note that $G^{0}_A$ is the species denoting that Gene A is in the OFF state and hence its copy-number is simply $(1-x_1)$. The interpretation for species $G^{0}_B$ and $G^{0}_C$ is similar. 
}
\label{tab:repress1}
\end{table}

For the CTMC model of this network, the state-space $\mathcal{E} = \mathcal{E}_b \times \N^6_0$ is irreducible, where $$\mathcal{E}_b = \{  (0,0,0), (0,0,1), (0,1,0), (0,1,1), (1,0,0), (1,0,1), (1,1,0), (1,1,1)\}$$ is the finite set which contains the dynamics of the copy-numbers $(x_1,x_2,x_3)$ of the three genes being in the ON state. The copy-numbers of all the mRNAs and proteins can take values in the whole set of non-negative integers $\N_0$. We apply sFSP on the 3-gene network with the finite truncated state-space $\mathcal{E}_i$ for sFSP iteration $i$ chosen as $\mathcal{E}_i = \mathcal{E}_{\bm n_i}$ (see \eqref{lattice_tensor_defn}) with
\begin{align*}
{\bm n_i} = (2,2,2,U_{m,i},U_{m,i},U_{m,i},U_{p,i},U_{p,i},U_{p,i}).
\end{align*}
Here $U_{m,i}$ and $U_{p,i}$ denote the strict upper-bounds for the copy-numbers of all the mRNAs and proteins respectively. The convergence factor $\gamma_i$ is computed for this example using $\beta_i =  \max \{ U_{m,i}, U_{p,i} \}$ in step 2 of Algorithm \ref{algo_sfsp}. Due to limitations posed by the qtt-toolbox and our computational hardware, we fix the acceptable threshold level $\epsilon$ (see step 4 of Algorithm \ref{algo_sfsp}) to be $10^{-2}$ instead of $10^{-10}$ used previously.

\begin{figure}[ht!]
\centering
\includegraphics[width=0.95\textwidth]{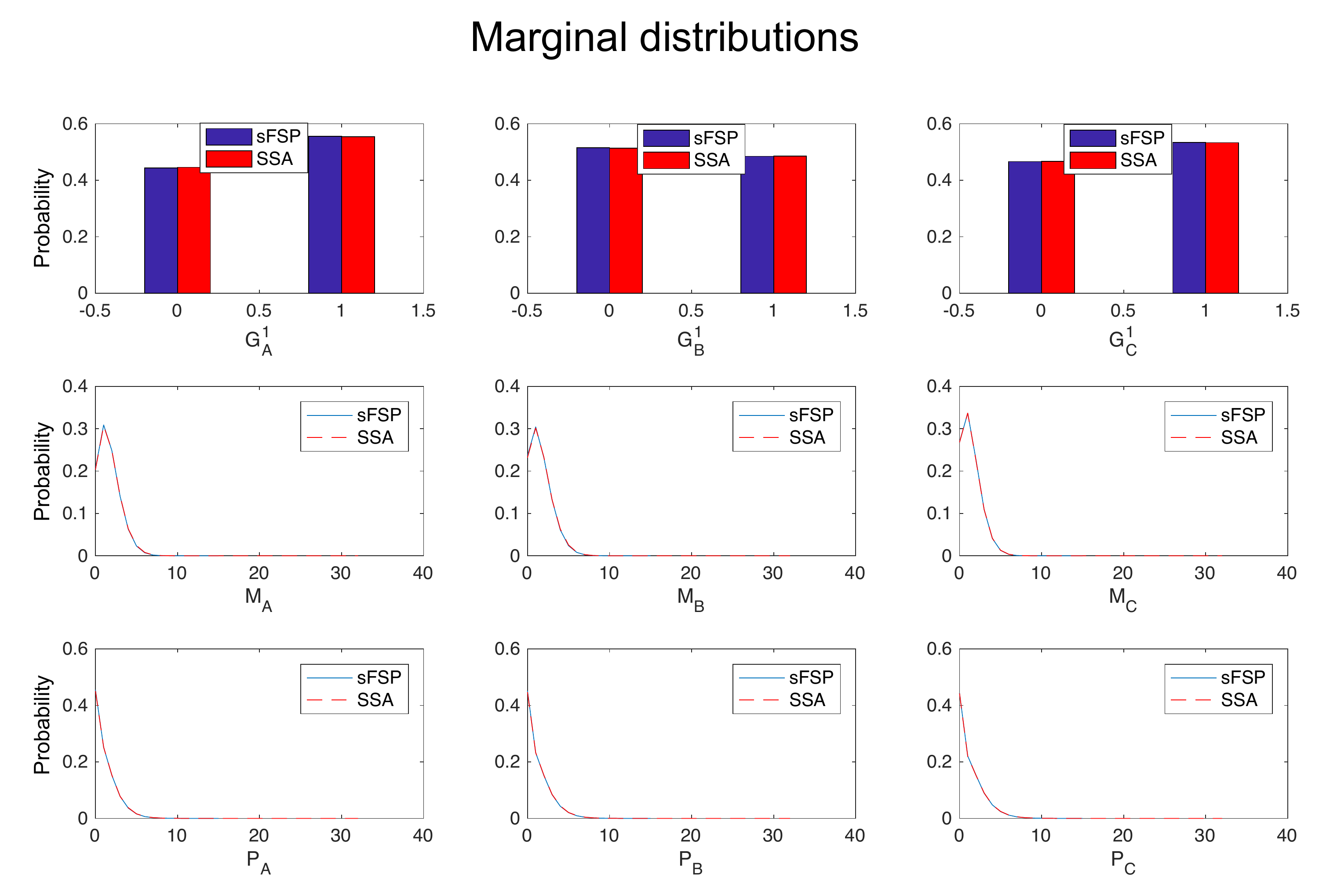}
\caption{In this figure the sFST estimated marginal stationary distributions for the all the $9$ species in the triple-repressor model are plotted and also compared with the distributions estimated with Gillespie's SSA.}
\label{fig:repress2} 
\end{figure}

The performance of sFSP on this triple-repressor network is summarized in Table \ref{table:sfsp_repress}, where for each iteration $i$, the upper-bounds ($U_{m,i}$ and $U_{p,i}$), the truncated state-space size ($| \mathcal{E}_i |$), the convergence factor $\gamma_i$ and the CPU times are provided. One can see that sFSP terminated in $5$ iterations and overall it required around 168 minutes of CPU time\footnote{All the computations for this QTT implementation of sFSP were performed on a Lenovo T440 machine with 1.6 GHz Intel i5-4200U processor with 8GB of RAM}. Note that the copy-numbers of all the species are relatively \emph{small} in this example. However due to the large number of species, the size of the final truncated state-space $\mathcal{E}_5$ is several times larger than the  truncated state-spaces encountered in the examples considered before, for the simple implementation of sFSP. To assess the accuracy of sFSP, we also estimate $\pi$ using $10^6$ CTMC trajectories simulated with SSA in the time-interval $[0,200]$. As in the previous examples, we plot the CPU times and the convergence factors at all the sFSP iterations in Figure \ref{fig:repress1}{\bf B}, the contour plots for the various joint stationary distributions estimated by sFSP in Figure \ref{fig:repress1}{\bf C}, and the estimated marginal stationary distributions for the all the $9$ species in Figure \ref{fig:repress2}. These marginal stationary distributions are also compared with the corresponding SSA-estimated marginal stationary distributions and one can see that the match is quite good.  

\begin{table}[h]
\begin{center}
\begin{tabular}{|c|c|c|c|c|c|c|}
\hline
Iteration & \multicolumn{2}{|c|}{Upper bounds}  & State-space size &
Convergence factor & CPU Time (minutes)   \\
\hline
$i$ & mRNAs $U_{m,i}$ & Proteins $U_{p,i}$ & $| \mathcal{E}_i |$& $\gamma_i = r^{(i)}_{ \textnormal{out} } U_{p,i}$ & t   \\
\hline
\hline
$1$ & $4$ & $4$ & $32,768$& $ 13.5607 $ & 3.66   \\
$2$ & $8$ & $4$ & $262,144$& $47.3662$ &  6.77   \\
$3$ & $8$ & $8$ & $2,097,152$& $2.4899$ &  27.67 \\
$4$ & $16$ & $8$ & $16,777,216$& $5.2869$ &  60.09   \\
$5$ & $16$ & $16$ & $134,217,728$& $0.0036 $ &   69.66   \\
\hline
\end{tabular}
\end{center}
\caption{Application of sFSP on the triple-repressor model. }
\label{table:sfsp_repress}
\end{table}

The SSA-based estimation with $10^6$ trajectories needed around 117 minutes of CPU time, based on a C++ implementation, which is slightly faster than sFSP (168 minutes). However we must note that even though this SSA-based estimation captures the marginal distributions very well (see Figure \ref{fig:repress2}), it is unable to capture the full stationary distribution because the state-space is high-dimensional and the size of the final truncated state-space $\mathcal{E}_5$ for sFSP suggests that the support of the true stationary distribution is much larger ($>$130 million) than the number of SSA samples ($1$ million) being used for the estimation. To illustrate this point, we compute the $\ell_1$ distance between the sFSP estimated stationary distribution $\bar{\pi}$ and the stationary distribution $\hat{\pi}$ estimated with $ 10^5, 10^6$ and $10^{7}$ SSA samples. The results are shown in Table \ref{table:ssacomp} along with the associated CPU times for generating the SSA samples. Notice that as the number of SSA samples increases, the $\ell_1$ distance $ \| \bar{\pi} - \hat{\pi}  \|_{\ell_1}$ decreases sharply, which strongly suggests that sFSP is an accurate approximation of the true stationary distribution $\pi$. However this $\ell_1$ distance is significant when $\hat{\pi}$ is estimated with $1$ million SSA samples, which implies that $\hat{\pi}$ is quite inaccurate. If we use $10^7$ SSA samples to estimate $\hat{\pi}$ then the accuracy improves but the total CPU time required is approximately 18 hours, that is $6.4$ times larger than the time needed for sFSP.

\begin{table}[h]
\begin{center}
\begin{tabular}{|c|c|c|}
\hline
No. of SSA samples &$ \| \bar{\pi} - \hat{\pi}  \|_{\ell_1}$ &  CPU Time \\
\hline
 $10^5$ & 0.5969 & 12 minutes\\
  $10^6$ & 0.2461 & 117 minutes \\
   $10^7$ & 0.091 & 1076 minutes\\
\hline
\end{tabular}
\end{center}
\caption{Comparison of the sFSP estimated stationary distribution $\bar{\pi}$ and the SSA estimated stationary distribution $\hat{\pi}$ for the triple-repressor model. Computed $\ell_1$ distance $ \| \bar{\pi} - \hat{\pi}  \|_{\ell_1} $ and CPU times to generate SSA samples are shown for three sample sizes $ 10^5, 10^6$ and $10^{7}$.}
\label{table:ssacomp}
\end{table}

\section{Conclusion} \label{sec:conclusion}

In this paper we presented a new method for estimating the stationary probability distributions of continuous-time Markov chain (CTMC) models of reaction networks based on suitable truncations of the CME. The method which we call the \emph{stationary Finite State Projection} (sFSP) algorithm is similar to the Finite State Projection (FSP) algorithm\cite{FSP}, with the crucial difference being that instead of introducing an absorbing state, we redirect all the outgoing transitions from the truncated state-space to a designated state within the truncated state-space (see Figure \ref{figure:lattice}{\bf C}). This simple modification creates a \emph{projected} CTMC over the truncated state-space, whose stationary distribution can be obtained by solving a finite linear-algebraic system. We provided theoretical arguments to establish that this stationary distribution estimated from the projected CTMC is unique, converges to the true stationary distribution as the truncated state-space expands to the full state-space and for any truncated state-space the error between the estimated stationary distribution and the true stationary distribution can be assessed by computing the overall rate of outgoing transitions at the estimated stationary distribution (see Theorem \ref{thm:mainresult}). These results form the basis of our sFSP method. We illustrated the efficiency and accuracy of this method using several examples. These examples indicated that sFSP can easily outperform the stochastic simulation-based approach for estimating the stationary distribution, both in terms of computational speed as well as accuracy. This is not unexpected, as stochastic simulations are expensive to perform over large time-intervals, and the stationary distribution they estimate suffers from statistical errors that can be significant in regions of the state-space where the probabilities are extremely low. These issues do not arise in sFSP and this makes it an appealing method for estimating stationary distributions of CTMCs representing reaction networks.

There are several ways to improve and extend sFSP. Like FSP, this method is iterative in nature and the number of iterations it requires to converge depends on the specifics of the implementation of sFSP. In this paper we discussed two such implementations. In the first implementation the state-space was explicitly enumerated with \emph{Cantor pairing functions} and then truncated in \emph{trapezoidal} shapes (see Section \ref{secsimplesfsp}), while the second implementation was based on the recently developed \emph{quantized tensor train} (QTT) version of CME where each state-space truncation is a \emph{hyper-rectangle} (see Section \ref{secqttsfsp}). Both these implementations will benefit from better state-space truncation schemes that adapt to the problem at hand. One way to do this would be to use Lyapunov function theory or use stationary moment bounds to construct optimal state-space truncations (see \cite{Verena}, \cite{kuntz2017rigorous} and \cite{GuptaPLOS}). Observe that unlike FSP which solves a linear system of ODEs, sFSP only requires solving a linear-algebraic system which is computationally much easier. Hence sFSP can handle a wider range of networks in comparison to FSP. Indeed with the QTT implementation, finding stationary distributions for problems with state truncations exceeding 100 million states was shown to be feasible. A possible approach for enhancing the feasibility of sFSP to even larger problems would be to integrate it with sparse grids and aggregation methods \cite{SparseGrid1}. Note that at the core of sFSP, is the problem of finding vectors in the one-dimensional null-spaces of large, but extremely sparse matrices (see Section \ref{sec:implementation1}). This sparsity and the structure of the matrices that arise, make this problem quite amenable to parallel-computing approaches \cite{tran1996direct}.

\section*{Acknowledgments}
The authors would like to thank Prof. Sean Meyn (University of Florida) and Prof. Brian Munsky (Colorado State University) for their helpful comments and suggestions.

\bibliographystyle{unsrt}

\begin{thebibliography}{10}

\bibitem{Arkin}
Harley~H. McAdams and Adam Arkin.
\newblock Stochastic mechanisms in gene expression.
\newblock {\em Proc. Natl. Acad. Sci., Biochemistry}, 94:814--819, 1997.

\bibitem{Elowitz}
Michael~B. Elowitz, Arnold~J. Levine, Eric~D. Siggia, and Peter~S. Swain.
\newblock Stochastic gene expression in a single cell.
\newblock {\em Science}, 297(5584):1183--1186, 2002.

\bibitem{DASurvey}
D.A. Anderson and T.G. Kurtz.
\newblock Continuous time {M}arkov chain models for chemical reaction networks.
\newblock In H.~Koeppl, G.~Setti, M.~di~Bernardo, and D.~Densmore, editors,
  {\em Design and Analysis of Biomolecular Circuits}. Springer-Verlag, 2011.

\bibitem{GP}
Daniel~T. Gillespie.
\newblock Exact stochastic simulation of coupled chemical reactions.
\newblock {\em The Journal of Physical Chemistry}, 81(25):2340--2361, 1977.

\bibitem{FSP}
B.~Munsky and M.~Khammash.
\newblock The finite state projection algorithm for the solution of the
  chemical master equation.
\newblock {\em Journal of Chemical Physics}, 124(4), 2006.

\bibitem{Krylov}
Shev MacNamara, Kevin Burrage, and Roger~B Sidje.
\newblock Multiscale modeling of chemical kinetics via the master equation.
\newblock {\em Multiscale Modeling \& Simulation}, 6(4):1146--1168, 2008.

\bibitem{Kazeev}
Vladimir Kazeev, Mustafa Khammash, Michael Nip, and Christoph Schwab.
\newblock Direct solution of the chemical master equation using quantized
  tensor trains.
\newblock {\em PLoS Comput Biol}, 10(3):e1003359, 03 2014.

\bibitem{SparseGrid1}
Markus Hegland, Andreas Hellander, and Per L{\"o}tstedt.
\newblock Sparse grids and hybrid methods for the chemical master equation.
\newblock {\em BIT Numerical Mathematics}, 48(2):265, 2008.

\bibitem{KurtzLLn2}
Thomas~G. Kurtz.
\newblock Strong approximation theorems for density dependent {M}arkov chains.
\newblock {\em Stochastic Processes Appl.}, 6(3):223--240, 1977/78.

\bibitem{Hellander}
Andreas Hellander and Per L{\"o}tstedt.
\newblock Hybrid method for the chemical master equation.
\newblock {\em Journal of Computational Physics}, 227(1):100--122, 2007.

\bibitem{Benni}
Benjamin Hepp, Ankit Gupta, and Mustafa Khammash.
\newblock Adaptive hybrid simulations for multiscale stochastic reaction
  networks.
\newblock {\em The Journal of chemical physics}, 142(3):034118, 2015.

\bibitem{Kazeev2015}
Vladimir Kazeev and Christoph Schwab.
\newblock Tensor approximation of stationary distributions of chemical reaction
  networks.
\newblock {\em SIAM Journal on Matrix Analysis and Applications},
  36(3):1221--1247, 2015.

\bibitem{ACKProd}
David~F. Anderson, Gheorghe Craciun, and Thomas~G. Kurtz.
\newblock Product-form stationary distributions for deficiency zero chemical
  reaction networks.
\newblock {\em Bulletin of Mathematical Biology}, 72(8):1947--1970, 2010.

\bibitem{Norris}
J.~R. Norris.
\newblock {\em Markov chains}, volume~2 of {\em Cambridge Series in Statistical
  and Probabilistic Mathematics}.
\newblock Cambridge University Press, Cambridge, 1998.
\newblock Reprint of 1997 original.

\bibitem{GuptaIrred}
Ankit Gupta and Mustafa Khammash.
\newblock A generic state-space decomposition method for analyzing stochastic
  biomolecular reaction networks.
\newblock {\em arXiv preprint arXiv:1505.06594}, 2017.

\bibitem{GuptaPLOS}
Ankit Gupta, Corentin Briat, and Mustafa Khammash.
\newblock A scalable computational framework for establishing long-term
  behavior of stochastic reaction networks.
\newblock {\em PLoS Comput Biol}, 10(6):e1003669, 06 2014.

\bibitem{Gauckler}
Ludwig Gauckler and Harry Yserentant.
\newblock Regularity and approximability of the solutions to the chemical
  master equation.
\newblock {\em ESAIM: Mathematical Modelling and Numerical Analysis},
  48(6):1757--1775, 2014.

\bibitem{Verena}
Tugrul Dayar, Holger Hermanns, David Spieler, and Verena Wolf.
\newblock Bounding the equilibrium distribution of markov population models.
\newblock {\em Numerical linear algebra with applications}, 18(6):931--946,
  2011.

\bibitem{Meyn}
Sean~P. Meyn and R.~L. Tweedie.
\newblock Stability of {M}arkovian processes. {III}. {F}oster-{L}yapunov
  criteria for continuous-time processes.
\newblock {\em Adv. in Appl. Probab.}, 25(3):518--548, 1993.

\bibitem{EK}
S.~N. Ethier and T.~G. Kurtz.
\newblock {\em Markov processes : {C}haracterization and {C}onvergence}.
\newblock Wiley Series in Probability and Mathematical Statistics: Probability
  and Mathematical Statistics. John Wiley \& Sons Inc., New York, 1986.

\bibitem{Kemeny}
John~G. Kemeny and J.~Laurie Snell.
\newblock {\em Finite {M}arkov chains}.
\newblock The University Series in Undergraduate Mathematics. D. Van Nostrand
  Co., Inc., Princeton, N.J.-Toronto-London-New York, 1960.

\bibitem{MeynandTweedieBook}
Sean Meyn and Richard~L. Tweedie.
\newblock {\em Markov chains and stochastic stability}.
\newblock Cambridge University Press, Cambridge, second edition, 2009.
\newblock With a prologue by Peter W. Glynn.

\bibitem{meyn1994computable}
Sean~P Meyn and Robert~L Tweedie.
\newblock Computable bounds for geometric convergence rates of markov chains.
\newblock {\em The Annals of Applied Probability}, pages 981--1011, 1994.

\bibitem{Hart}
Andrew~G Hart and Richard~L Tweedie.
\newblock Convergence of invariant measures of truncation approximations to
  markov processes.
\newblock {\em Applied Mathematics}, 3(12):2205, 2012.

\bibitem{CantorPairing}
Meri Lisi.
\newblock Some remarks on the cantor pairing function.
\newblock {\em Le Matematiche}, 62(1):55--65, 2007.

\bibitem{Arnoldi}
Richard~B Lehoucq and Danny~C Sorensen.
\newblock Deflation techniques for an implicitly restarted arnoldi iteration.
\newblock {\em SIAM Journal on Matrix Analysis and Applications},
  17(4):789--821, 1996.

\bibitem{MO}
Mukund Thattai and Alexander van Oudenaarden.
\newblock Intrinsic noise in gene regulatory networks.
\newblock {\em Proceedings of the National Academy of Sciences},
  98(15):8614--8619, 2001.

\bibitem{Gardner}
Timothy~S. Gardner, Charles~R. Cantor, and James~J. Collins.
\newblock Construction of a genetic toggle switch in escherichia coli.
\newblock {\em Nature}, 403(6767):339--342, 2000.

\bibitem{Keener}
James~P. Keener and James Sneyd.
\newblock {\em Mathematical physiology}, volume Interdisciplinary applied
  mathematics.
\newblock Springer, 2009.

\bibitem{fox2016finite}
Zachary Fox, Gregor Neuert, and Brian Munsky.
\newblock Finite state projection based bounds to compare chemical master
  equation models using single-cell data.
\newblock {\em The Journal of Chemical Physics}, 145(7):074101, 2016.

\bibitem{oseledets2011tensor}
Ivan~V Oseledets.
\newblock Tensor-train decomposition.
\newblock {\em SIAM Journal on Scientific Computing}, 33(5):2295--2317, 2011.

\bibitem{oseledets2010approximation}
Ivan~V Oseledets.
\newblock Approximation of 2\^{}d$\backslash$times2\^{}d matrices using tensor
  decomposition.
\newblock {\em SIAM Journal on Matrix Analysis and Applications},
  31(4):2130--2145, 2010.

\bibitem{oseledets2009approximation}
IV~Oseledets.
\newblock Approximation of matrices with logarithmic number of parameters.
\newblock In {\em Doklady Mathematics}, volume~80, pages 653--654. Springer,
  2009.

\bibitem{ipsen1997computing}
Ilse~CF Ipsen.
\newblock Computing an eigenvector with inverse iteration.
\newblock {\em SIAM review}, 39(2):254--291, 1997.

\bibitem{rakhuba2016calculating}
Maxim Rakhuba and Ivan Oseledets.
\newblock Calculating vibrational spectra of molecules using tensor train
  decomposition.
\newblock {\em The Journal of Chemical Physics}, 145(12):124101, 2016.

\bibitem{qtttoolbox}
I~Oseledets, S~Dolgov, V~Kazeev, O~Lebedeva, and T~Mach.
\newblock {QTT-Toolbox}.
\newblock \url{http://oseledets.github.io/software/}.

\bibitem{elowitz2000synthetic}
Michael~B Elowitz and Stanislas Leibler.
\newblock A synthetic oscillatory network of transcriptional regulators.
\newblock {\em Nature}, 403(6767):335--338, 2000.

\bibitem{kuntz2017rigorous}
Juan Kuntz, Philipp Thomas, Guy-Bart Stan, and Mauricio Barahona.
\newblock Rigorous bounds on the stationary distributions of the chemical
  master equation via mathematical programming.
\newblock {\em arXiv preprint arXiv:1702.05468}, 2017.

\bibitem{tran1996direct}
TM~Tran, R~Gruber, K~Appert, and S~Wuthrich.
\newblock A direct parallel sparse matrix solver.
\newblock {\em Computer physics communications}, 96(2-3):118--128, 1996.

\end{thebibliography}

\end{document}